\documentclass[final,3p,times]{elsarticle}
\usepackage[latin9]{inputenc}
\usepackage{array}
\usepackage{verbatim}
\usepackage{bm}
\usepackage{multirow}
\usepackage{amsmath}
\usepackage{amssymb}
\usepackage{graphicx}
\usepackage{esint}
\usepackage[unicode=true,
 bookmarks=false,
 breaklinks=false,pdfborder={0 0 1},colorlinks=false]
 {hyperref}

\usepackage{tikz}
\usetikzlibrary{shapes.geometric, arrows, positioning, matrix}

\makeatletter

\usepackage{tabularx}\usepackage{subfigure}\usepackage{subfigure}\usepackage{color}\usepackage{textcomp}\usepackage{latexsym}

\usepackage{setspace}

\usepackage{bm}

\journal{ }

\makeatother

\begin{document}

\title{Discontinuity-resolving shock-capturing schemes on unstructured grids}

\author[ad1,ad4]{Lidong Cheng}

\author[ad2]{Xi Deng\corref{cor1}}

\author[ad3]{Bin Xie\corref{cor1}}

\author[ad1]{Yi Jiang\corref{cor1}}

\author[ad4]{Feng Xiao\corref{cor1}}

\address[ad1]{School of Aerospace Engineering, Beijing Institute of Technology, Beijing, 10-0081, China}

\address[ad2]{Department of Aeronautics,Imperial College London,SW7 2AZ, United Kingdom.}

\address[ad3]{School of Naval Architecture, Department of Ocean and Civil Engineering, Shanghai Jiaotong University, Shanghai, 200240, China.}

\address[ad4]{School of Engineering, Department of Mechanical Engineering, Tokyo Institute of Technology, Tokyo, 152-8550, Japan}

\cortext[cor1]{Corresponding authors: Dr. X. Deng (Email:deng.xi98@gmail.com), Dr. B. Xie (Email:xie.b.aa@sjtu.edu.cn), Dr. Y. Jiang (Email:jy2818@163.com), Dr. F. Xiao (Email: xiao.f.aa@m.titech.ac.jp)}


\begin{abstract}
Solving compressible flows containing discontinuities remains a major challenge for numerical methods especially on unstructured grids. Thus in this work, we make contributions to shock capturing schemes on unstructured grids with aim of resolving discontinuities with low numerical dissipation. Different from conventional shock capturing schemes which only use polynomials as interpolation functions on unstructured grids, the proposed scheme employs the linear polynomial as well as non-polynomial as reconstruction candidates. For linear polynomial, the second order MUSCL (Monotone Upstream-centered Schemes for Conservation law) scheme with the MLP (Multi-dimensional Limiting Process) slope limiter is adopted. The multi-dimensional THINC (Tangent of Hyperbola for INterface Capturing) function with quadratic surface representation and Gaussian quadrature, so-called THINC/QQ, is used as the non-polynomial reconstruction candidate. With these reconstruction candidates, a multi-stage boundary variation diminishing (BVD) algorithm which aims to minimize numerical dissipation is designed on unstructured grids to select the final reconstruction function. The resulted shock capturing scheme is named as MUSCL-THINC/QQ-BVD. The performance of the proposed scheme is demonstrated through solving compressible single-phase and multi-phase problems where the discontinuity is the typical flow structure. The numerical results show that the proposed scheme is capable of capturing sharp discontinuous profiles without numerical oscillations as well as resolving vortices associated with Kelvin-Helmholtz instabilities along shear layers and material interfaces. In comparison with schemes only replying on high order polynomials, the proposed scheme shows significant improvement of resolution across discontinuities. Thus, this work provides an accurate and robust shock-capturing scheme to resolve discontinuities in compressible flows.  

\end{abstract}

\begin{keyword}
compressible flows \sep discontinuities \sep low-dissipation \sep BVD algorithm \sep unstructured grids.
\end{keyword}
\maketitle

\section{Introduction}
Compressible flows containing discontinuities are widely found in transonic, supersonic and hypersonic flows, as well as flows involving multi-phases and multi-species. Sharp flow features such as shock waves, contact discontinuities, moving material interfaces, strong species gradients and shear layers become typical in these flows. Since in many cases compressible flows are too complicate to be analyzed by theoretical analysis or experimental approaches, the numerical simulation becomes an effective alternative to provide rich flow information for investigating the fundamental mechanisms. Meanwhile, industrial applications usually involve complex geometrical boundaries thus numerical solver on unstructured meshes are preferable.

Over recent decades, a number of high order schemes have been proposed on unstructured grids to provide high resolution solution for compressible flows. In the finite volume framework, a well-known approach called $k$-exact least-square method has been proposed and developed in \cite{k1,k2,k3}. In this approach, a stencil consisting of the target cell and its neighbors is employed to construct a polynomial of degree $k$. For higher degree of polynomials, however, the stencil has to be extended to include more neighbor cells, which increases the complexity for the algorithm and parallelism. Different from conventional finite volume method, high order Discontinuous Galerkin (DG) \cite{dg1,dg2}, Flux Reconstruction (FR) \cite{fr1,fr2,fr3,fr4} and Spectral Difference (SD) methods \cite{sd1,sd2} realize high-order reconstruction by using locally defined degrees of freedom (DOFs). These methods have received particular attention in recent years because of their superior convergence property as well as the compact stencil. 

Although these schemes based on high order polynomial reconstructions show superiority in simulating smooth flow features such as acoustic waves and turbulence, high order schemes generally face challenges to obtain accurate and stable solutions around discontinuities. Special techniques such as limiting projection or artificial viscosity must be designed to prevent Gibbs phenomenon and its associated spurious numerical oscillations arising from high order interpolations. Especially, solving discontinuous flow features on unstructured grids imposes additional difficulty. Based on the idea of TVD (Total Variation Diminishing) schemes \cite{harten1983high}, multi-dimensional slope limiting processes on unstructured grids have been proposed in \cite{tvdm1,tvdm2} and improved recently in \cite{mlp1,park2012multi}. In order to reduce the numerical dissipation of TVD schemes, WENO schemes (Weighted Essentially Non-Oscillatory) have been extended to unstructured grids, for example \cite{friedrich1998weighted, haselbacher2005weno, wolf2007high,titarev2010weno,tsoutsanis2011weno}, to cite but a few. The general idea of WENO schemes is to construct a weighted average of the polynomial approximations over all candidate stencils. However, it is not a trivial work to construct efficient WENO scheme on unstructured grids since dealing with wide stencils and choosing admissible stencil cells increases the algorithmic complexity. The difficulty of solving discontinuities also exists and is more severe for high order local reconstruction such as DG and FR schemes. Although several strategies such as artificial viscosity \cite{av1,av2} and subcell finite volume formulation \cite{sb1,sb2} have been proposed and improved, solving discontinuities accurately and robustly remains a challenge for high order local reconstruction schemes. Moreover, in spite of the efforts aforementioned, limiting processes or artificial viscosity methods usually introduce excessive numerical dissipation which continuously smears and blurs flow structures. Especially, the resolution of discontinuous flow features such as contact surfaces, shear waves, reaction fronts and material interfaces may evolve from bad to worse due to the limiting process. 

Realizing that the polynomial-based reconstruction may not be a proper choice when the solution includes both smooth and discontinuous flow structures, the work \cite{sun2016boundary} proposed a novel algorithm called boundary variation diminishing (BVD) which selects non-polynomial-based reconstruction THINC (Tangent of Hyperbola for INterface Capturing) scheme \cite{xiao2005simple} to solve discontinuous flow structures while high order polynomial-based WENO scheme \cite{jiang1996efficient} for smooth flow regions. The proposed methodology has significantly reduces the numerical dissipation across discontinuities. Following the work \cite{sun2016boundary}, the BVD algorithm has been applied for more challenging problems involving stiff source terms and material interfaces in the work of \cite{deng2018limiter,deng2018high}. More recently, the works \cite{deng2019fifth,deng2020constructing} devise higher order shock capturing schemes which retain the high resolution property in smooth region through the BVD algorithm.
Although being mainly practiced on structured grids, the above works show the BVD algorithm provides an alternative method to design accurate and robust shock capturing schemes.

In this work, we make efforts to extend the BVD algorithm on unstructured grids and devise new shock capturing schemes which are capable of resolving discontinuities with high resolutions. The proposed scheme employs the linear polynomial as well as non-polynomial as reconstruction candidates. For linear polynomial, the second order MUSCL (Monotone Upstream-centered Schemes for Conservation law) scheme with the MLP (Multi-dimensional Limiting Process) slope limiter \cite{mlp1,park2012multi} is adopted. The multi-dimensional THINC (Tangent of Hyperbola for INterface Capturing) function with quadratic surface representation and Gaussian quadrature \cite{xie2017toward,xie2019high}, so-called THINC/QQ, is used as the non-polynomial reconstruction candidate. With above reconstruction candidates, a multi-stage BVD algorithm is devised to select the final reconstruction function. The resulted shock capturing scheme is named as MUSCL-THINC/QQ-BVD. The performance of MUSCL-THINC/QQ-BVD scheme is demonstrated through solving compressible single-phase and multi-phase problems. The numerical results show that the proposed scheme is able to capture sharp discontinuous profiles without numerical oscillation. Also, it's able to resolve vortices associated with Kelvin-Helmholtz instabilities along the shear plane and material interface. Thus the proposed MUSCL-THINC/QQ-BVD scheme   significantly improves the resolution across discontinuities in comparison with schemes only replying on high order polynomials. The proposed scheme is expected to serve as an accurate and robust shock capturing scheme for problems where the discontinuity is the typical flow structure.

The rest of this paper is organized as follows. Mathematical models for numerical tests are introduced in section 2. Section 3 is a brief introduction to the MUSCL scheme and the global THINC/QQ scheme, followed by details of our BVD algorithm and two new BVD schemes. Numerical results and discussion are presented in section 4 and some concluding remarks in section 5.

\section{Mathematical models}

\subsection{Governing equations}

 In this work, our numerical schemes are tested by linear advection problems, inviscid single-phase and two-component compressible flows. Only two dimensional problems are discussed here. A general form of conservation laws can be written as:
 \begin{equation} \label{eq:consLaw}
    \frac{\partial \bm{U}}{\partial t} + \frac{\partial \bm{F}(\bm{U})}{\partial x} + \frac{\partial \bm{G}(\bm{U})}{\partial y} = \bm{S}
\end{equation}
\begin{itemize}



\item The Euler equation

Inviscid single-phase compressible flows are modeled by the Euler equation. It consists of equations for conservation of mass, momentum and energy respectively.
\begin{equation} \label{eq:euler}
    \bm{U}   =\left( \begin{array}{cc} \rho   \\ \rho u    \\ \rho v     \\ E \end{array} \right),
    \bm{F}(\bm{U})=\left( \begin{array}{cc} \rho u \\ \rho uu+p \\ \rho uv    \\ u(E+p) \end{array} \right),
    \bm{G}(\bm{U})=\left( \begin{array}{cc} \rho v \\ \rho vu   \\ \rho vv +p \\ v(E+p) \end{array} \right),
    \bm{S}   =\left( \begin{array}{cc} 0      \\  0        \\  0         \\ 0 \end{array} \right)
\end{equation}
where $\rho$ is density, $p$ is pressure field, and $E$ is total energy.

\item The five-equation model

Inviscid two-phases compressible flows under mechanical equilibrium are modeled by the five-equation model developed in \cite{allaire2002five}. It assumes that interface cells containing two kinds of fluids are in equilibrium of pressure. Governing equations consist of two mass conservation law, two momentum equations, one energy equation and an equation for the transportation of volume fraction.
\begin{equation} \label{eq:fiveEq}
    \bm{U}=\left( \begin{array}{cc} \alpha_1 \\ \rho_1 \alpha_1 \\ \rho_2 \alpha_2 \\ \rho u \\ \rho v \\ E \end{array} \right),
    \bm{F}(\bm{U})=\left( \begin{array}{cc} u\alpha_1 \\ \rho_1 \alpha_1 u \\ \rho_2 \alpha_2 u \\ \rho uu+p \\ \rho uv \\ u(E+p) \end{array} \right),
    \bm{G}(\bm{U})=\left( \begin{array}{cc} v\alpha_1 \\ \rho_1 \alpha_1 v \\ \rho_2 \alpha_2 v \\ \rho vu \\ \rho vv +p \\ v(E+p) \end{array} \right),
    \bm{S}=\left( \begin{array}{cc} \alpha_1 \nabla \cdot \bm{V} \\ 0 \\ 0 \\  0 \\  0 \\ 0 \end{array} \right)
\end{equation}
where $\alpha_k \in \left[ 0,1 \right]$ and $\rho_k$ is the volume fraction and density of the $k$th $(k=1,2)$ fluid. $\bm{V}=(u,v)$ is the velocity field.
\end{itemize}

\subsection{The closure strategy}

To close the Euler equation and the five-equation model, fluids are assumed to satisfy the following ideal gas law:
\begin{equation} \label{eq:eos}
    p=\rho e (\gamma - 1)
\end{equation}
where $e$ is the internal energy, and $\gamma$ is the ratio of the specific heats.

For two-component flows, conservative constraints lead to the following mixing formula of volume fraction, density and internal energy:
\begin{equation} \label{eq:mixLaw}
    \begin{array}{cc}
    \alpha_1 + \alpha_2 = 1  \\
    \alpha_1 \rho_1 + \alpha_2 \rho_2 = \rho \\
    \alpha_1 \rho_1 e_1 + \alpha_2 \rho_2 e_2 = \rho e 
    \end{array}
\end{equation}
As derived in \cite{shyue2001fluid}, the mixed ratio of the specific heats can be calculated as
\begin{equation} \label{eq:mixPress}
      \frac{1}{\gamma-1} = \frac{\alpha_1}{\gamma_1-1} + \frac{\alpha_2}{\gamma_2-1}.
\end{equation}

\section{Numerical methods}

\subsection{Computational grids}
Two dimensional computational domains are divided into non-overlapping triangular or quadrilateral elements $\Omega_i(i=0,1,2,...,N)$. Vertices and edges are denoted by $\vartheta_{ik}(k=1,2,...,K)$ and $\Gamma_{ij}(j=1,2,...,J)$, where $K=J=3$ for triangular meshes and $K=J=4$ for quadrilateral meshes. The cell center is denoted by $\vartheta_{ic}(x_{ic},y_{ic})$ . We define the area of element $\Omega_i$ as $\left| \Omega_i \right|$, length and unit normal vector of edge $\Gamma_{ij}$ as $\left| \Gamma_{ij} \right|$ and $\bm{n}_{ij}=(n_{ijx},n_{ijy})$.
\begin{figure}
    \centering
    \includegraphics[width=0.75\textwidth]{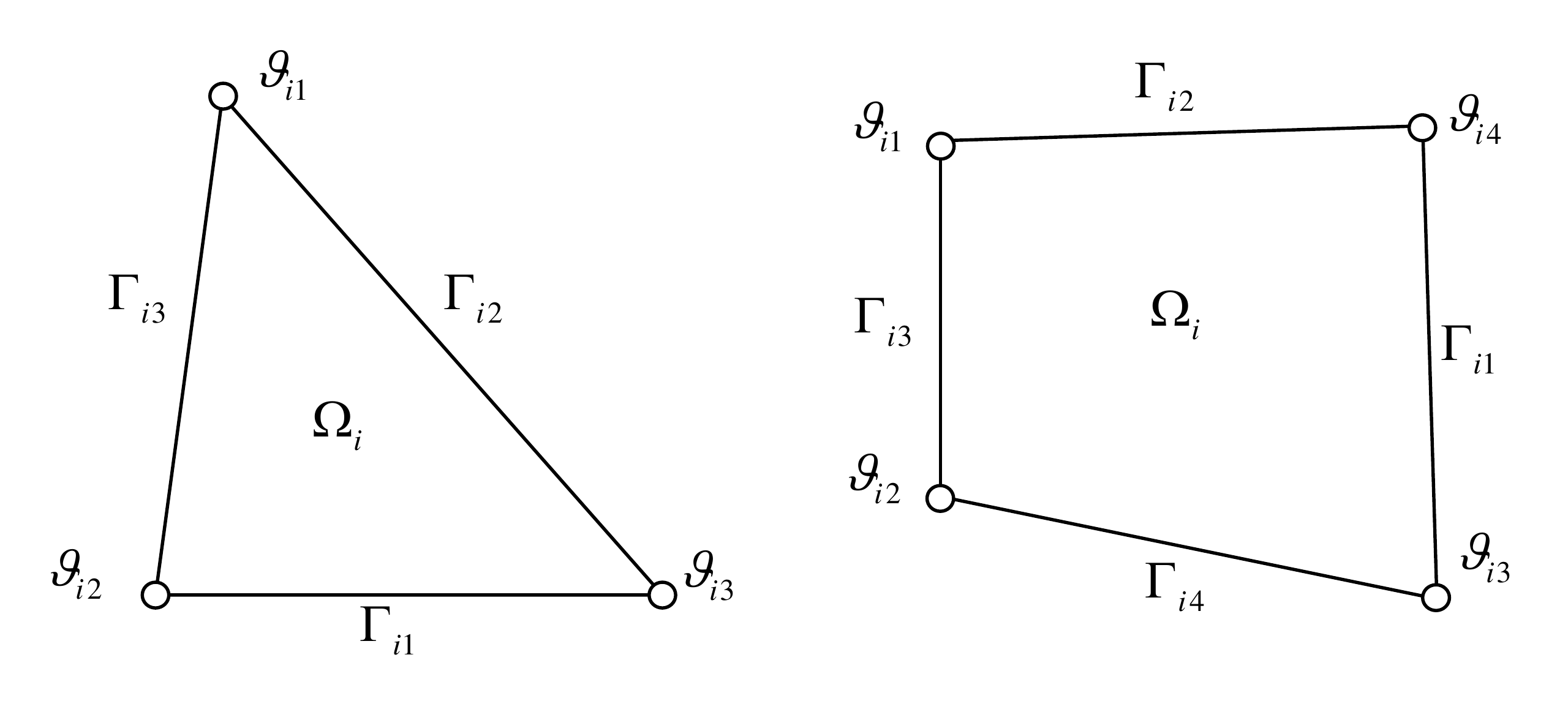}
    \caption{Two dimensional elements}
    \label{fig:element}
\end{figure}

\subsection{A Godunov-type finite volume method}
The first step is to integrate Eq. (\ref{eq:consLaw}) over a finite volume element $\Omega_i$ yielding the following semi-discrete form for cell-average values.
\begin{equation} \label{eq:integConsLaw}
    \frac{d \overline{\bm{U}}_i}{dt} = \Re (\overline{\bm{U}}_i) 
    = -\frac{1}{\left| \Omega_i \right|} \oint_{\partial \Omega_i} \bm{F}_n(\bm{U})d \Gamma + \frac{1}{\left| \Omega_i \right|}\oint_{\Omega_i}\bm{S}d\Omega
\end{equation}
where $\bm{F}_n(\bm{U})=\bm{F}(\bm{U})n_{ijx}+\bm{G}(\bm{U})n_{ijy}$. Curve integration along boundaries of the element can be calculated by the summation of integration along each edge.
\begin{equation} \label{eq:integFlux}
    \oint_{\partial \Omega_i} \bm{F}_n(\bm{U})d \Gamma 
    = \sum_{j=1}^{J} \int_{\Gamma_{ij}} \bm{F}_n(\bm{U})d \Gamma
    \approx \sum_{j=1}^{J} \bm{F}_{nij}(\bm{U}) \left| \Gamma_{ij} \right|
    = \sum_{j=1}^{J} \bm{F}_{nij}(\bm{U}_{ij}^+,\bm{U}_{ij}^-) \left| \Gamma_{ij} \right|
\end{equation}
Here, $\bm{U}_{ij}^{\pm}$ are the right and left states at the edge $\Gamma_{ij}$. $\bm{F}_{nij}(\bm{U}_{ij}^+,\bm{U}_{ij}^-)$ is the numerical flux which can be calculated by Riemann solvers projected on the normal direction.


Finally, we can update cell-average values with the following third-order SSP Runge-Kutta \cite{gottlieb2001strong} method.
\begin{equation}
    \begin{array}{cc}
        \overline{\bm{U}}_i^{\ast} = \overline{\bm{U}}_i(t) + \triangle t \Re(\overline{\bm{U}}_i(t)) \\
        \overline{\bm{U}}_i^{\ast\ast} = \frac{3}{4} \overline{\bm{U}}_i(t) + \frac{1}{4} \overline{\bm{U}}_i^{\ast} + \frac{1}{4} \triangle t \Re(\overline{\bm{U}}_i^{\ast}) \\
        \overline{\bm{U}}_i(t+\triangle t) = \frac{1}{3} \overline{\bm{U}}_i(t) + \frac{2}{3} \overline{\bm{U}}_i^{\ast} + \frac{2}{3} \triangle t \Re(\overline{\bm{U}}_i^{\ast\ast})
    \end{array}
\end{equation}

\subsection{Reconstruction schemes}

In this subsection, we propose two MUSCL-THINC/QQ-BVD schemes to reconstruct the left and right states $\bm{U}_{ij}^{\pm}$ from cell-averaged value $\overline{\bm{U}}_i$. A MUSCL scheme with the MLP-u2 limiter and the global THINC/QQ schemes with different steepness are implemented as reconstruction candidates. A BVD algorithm is then devised to select final reconstruction function from candidates. To avoid numerical oscillation, we choose primitive variables as reconstructed variables \cite{johnsen2006implementation,johnsen20118665}, say, $(\rho,u,v,p)$ for the Euler equation and $(\alpha_1, \alpha_1\rho_1,\alpha_2\rho_2,u,v,p)$ for the five-equation model. We denote a single reconstructed variable by $q$ to simplify the introduction.

\subsubsection{The MUSCL scheme}

The basic idea of a MUSCL-type scheme is to reconstruct variables in a certain cell by linear distributions. We use it as a candidate scheme in BVD schemes to approximate smooth solutions. For two dimensional unstructured grids, it can be written as,
\begin{equation} \label{eq:muscl}
    q_{i}^{m}(x,y) = \Bar{q}_i + \phi_i \left( q_{xi}(x-x_{ic}) + q_{yi}(y-y_{ic}) \right)
\end{equation}
where $\Bar{q}_i$ is the cell-average value, $\phi_i$ is a slope limiter to keep monotonicity and suppress numerical oscillation, $\left( q_{xi},q_{yi} \right)$ is the cell-averaged gradient determined from least-square method.

We use the so-called MLP-u2 limiter in \cite{park2012multi} as the slope limiter for the MUSCL scheme.
\begin{displaymath}
    \phi_i = \mathop{\min}\limits_{k=1}^{K}\left\{ \begin{array}{ll}
             \Phi(R_{ik}) & \textrm{if } \left( \overline{\nabla q} \right)_i \cdot \bm{r}_{ik} \neq 0 \\
             1 & \textrm{otherwise}
            \end{array}
            \right.
\end{displaymath}
where $\bm{r}_{ik}(k=1,2,\dots,K)$ is the vector from $\vartheta_{ic}$ to vertex $\vartheta_{ik}$, and $R_{ik}$ is the ratio of the maximum or minimum allowable variation to the estimated variation at $\vartheta_{ik}$.
\begin{displaymath}
    R_{ik} = \max \left( 
            \frac{\Bar{q}_{ik}^{min}-\Bar{q}_i}{\left( \overline{\nabla q} \right)_i \cdot \bm{r}_{ik}}, 
            \frac{\Bar{q}_{ik}^{max}-\Bar{q}_i}{\left( \overline{\nabla q} \right)_i \cdot \bm{r}_{ik}}
        \right)
\end{displaymath}
Here, $\Bar{q}_{ik}^{min}$ and $\Bar{q}_{ik}^{max}$ are minimum and maximum cell-average values of cells around $\vartheta_{ik}$. For the MLP-u2 limiter,
\begin{displaymath}
    \Phi(R_{ik})=\frac{R_{ik}^2 + 2R_ik + \epsilon}{R_{ik}^2 + R_{ik} + 2.0 + \epsilon}
\end{displaymath}
where $\epsilon$ is a small positive number to distinguish a near smooth region from a fluctuating one. We take $\epsilon = 1.0\times10^{-15}$.

\subsubsection{The THINC/QQ scheme}

For unstructured grids, we use multi-dimensional THINC/QQ (THINC method with Quadratic surface representation and Gaussian Quadrature) scheme \cite{xie2019high, xie2017toward} as another candidate in BVD schemes. As shown in \cite{xie2019high, xie2017toward}, THINC/QQ is able to achieve high accuracy discontinuity representation by accounting of geometrical information such as normal direction and curvature of the discontinuity. 
Here we use THINC/QQ in global coordinates, which states
\begin{equation} \label{eq:thincqq}
    q_{i}^{l}(x,y) = \Bar{q}_i^{min} + \frac{\Bar{q}_i^{max} - \Bar{q}_i^{min}}{2} 
                \left( 1+\tanh \left( \frac{\beta}{H_i} \left( P_i(x,y) + d_i \right) \right) \right)
\end{equation}
where $\Bar{q}_i^{min}$ and $\Bar{q}_i^{max}$ is the maximum and minimum cell-average values of cells sharing vertices with cell $\Omega_i$.
\begin{displaymath}
    \Bar{q}_i^{max} = \mathop{max}\limits_{k=1}^K \left\{ \Bar{q}_{ik}^{max} \right\},
    \Bar{q}_i^{min} = \mathop{min}\limits_{k=1}^K \left\{ \Bar{q}_{ik}^{min} \right\}
\end{displaymath}
$\beta$ is a parameter to control the steepness. $H_i$ is the hydraulic diameter of $\Omega_i$. 
\begin{displaymath}
H_i = \frac{4 \left| \Omega_i \right|}{\sum_{j=1}^J \left| \Gamma_{ij} \right|} 
\end{displaymath}
$P_i(x,y) + d_i$ is a full quadratic polynomial including geometrical information of the reconstruction as
\begin{displaymath}
    P_i(x,y) = a_{20}(x-x_{ic})^2 + a_{11}(x-x_{ic})(y-y_{ic}) + a_{02}(y-y_{ic})^2 + a_{10}(x-x_{ic}) + a_{01}(y-y_{ic})
\end{displaymath}
Coefficients $a_{st}(0<s+t \le 2)$ can be calculated using least square method. The only unknown $d_i$ is determined from the conservation condition
\begin{displaymath}
    \frac{1}{\left| \Omega_i \right|} \oint_{\Omega_i} q_i(x,y)dxdy = \Bar{q}_i.
\end{displaymath}
This integration is calculated with Gaussian quadrature. For more details, please refer to  \cite{xie2019high,xie2017toward}.



\subsubsection{The BVD algorithm}

Generally, a numerical flux can be expressed in the following canonical formulation \cite{sun2016boundary,harten1983upstream}:
\begin{equation} \label{eq:numFlux2}
    \bm{F}(\bm{U}_L,\bm{U}_R)=\frac{1}{2}\left( \bm{F}(\bm{U}_L) + \bm{F}(\bm{U}_R) \right) - \Tilde{d}(\bm{U}_R-\bm{U}_L),
\end{equation}
$\Tilde{d}$ is a matrix function of $\bm{U}_R$ and $\bm{U}_L$. The last term of Eq.(\ref{eq:numFlux2}) can be interpreted as a diffusion term. The BVD algorithm is designed to select reconstruction function in target of minimizing the dissipation term. Choosing a final reconstruction with smaller boundary variation tends to preserve solution properties \cite{tann2019solution} and reduce numerical diffusion of the scheme.

\begin{figure}
    \centering
    \includegraphics[width=0.5\textwidth]{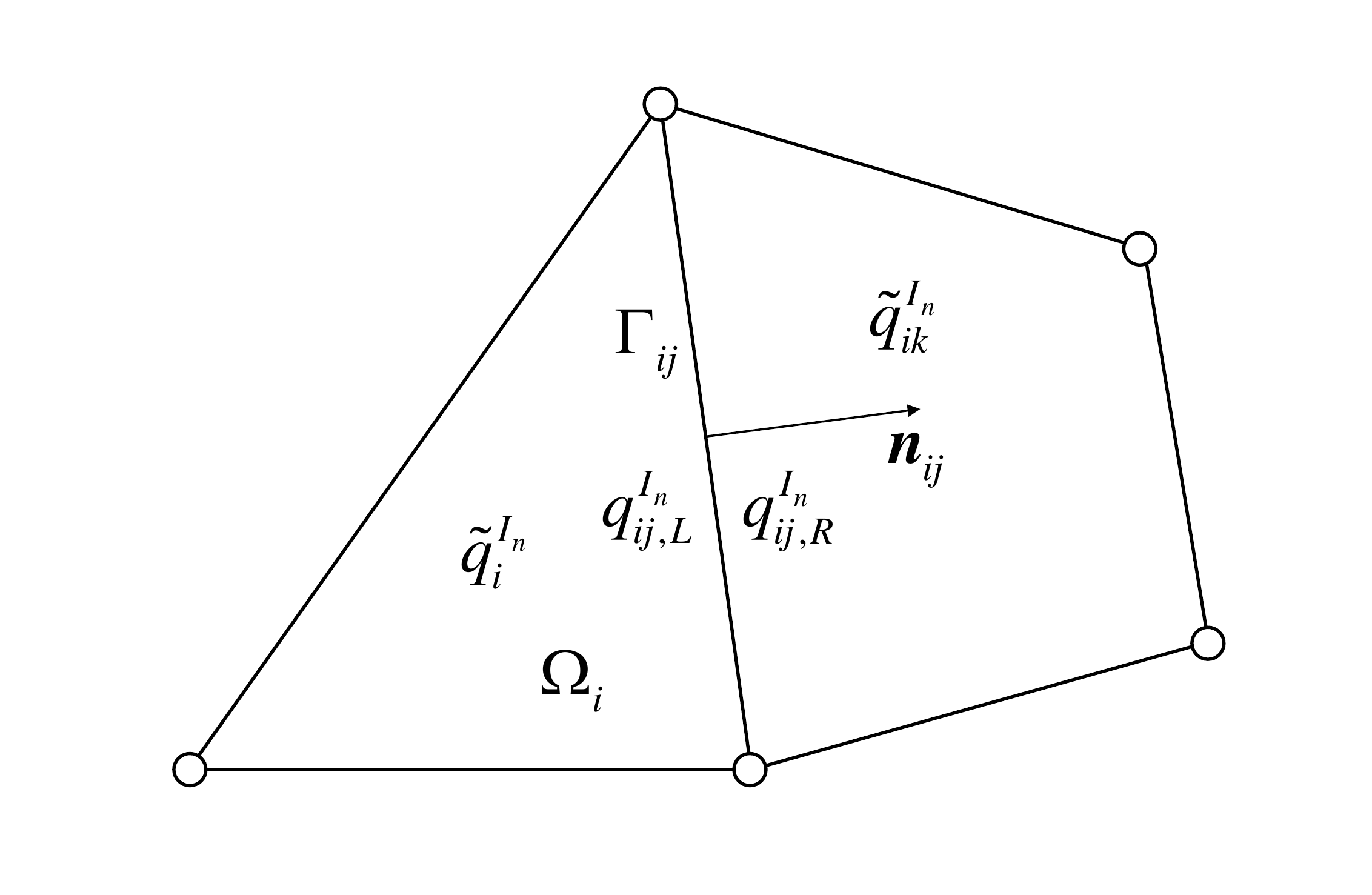}
    \caption{The left and right states of a cell face after reconstruction of candidate schemes}
    \label{fig:reconstruction}
\end{figure}



By assuming the target cell $i$ and its immediate neighbors are implemented with the same reconstruction scheme $I_b$ from the $B$ candidate schemes $(I_1,I_2,\cdots,I_B)$,  
we define the BV at cell edge $\Gamma_{ij}$ as
\begin{displaymath}
BV_{ij}^{I_b} = \left|q_{ij,L}^{I_b} - q_{ij,R}^{I_b} \right|\left|\Gamma_{ij}\right|.
\end{displaymath}
Then the total boundary variation ($TBV$) of cell $\Omega_i$ is defined as
\begin{displaymath}
TBV_i^{I_b} = \sum_{j=1}^J BV_{ij}^{I_b}.
\end{displaymath}
Finally, the BVD algorithm selects the reconstruction scheme $I_b$ to reconstruct $q$ in cell $\Omega_i$ if $I_b$ satisfies the following condition 
\begin{displaymath}
TBV_i^{I_b} = min(TBV_i^{I_1},TBV_i^{I_2},\cdots,TBV_i^{I_B}).
\end{displaymath}
As shown later in numerical results, the proposed BVD algorithm is able to select jump-like THINC function for discontinuities thus to significantly improve resolution.

\begin{figure}
    \centering
    \includegraphics[width=0.8\textwidth]{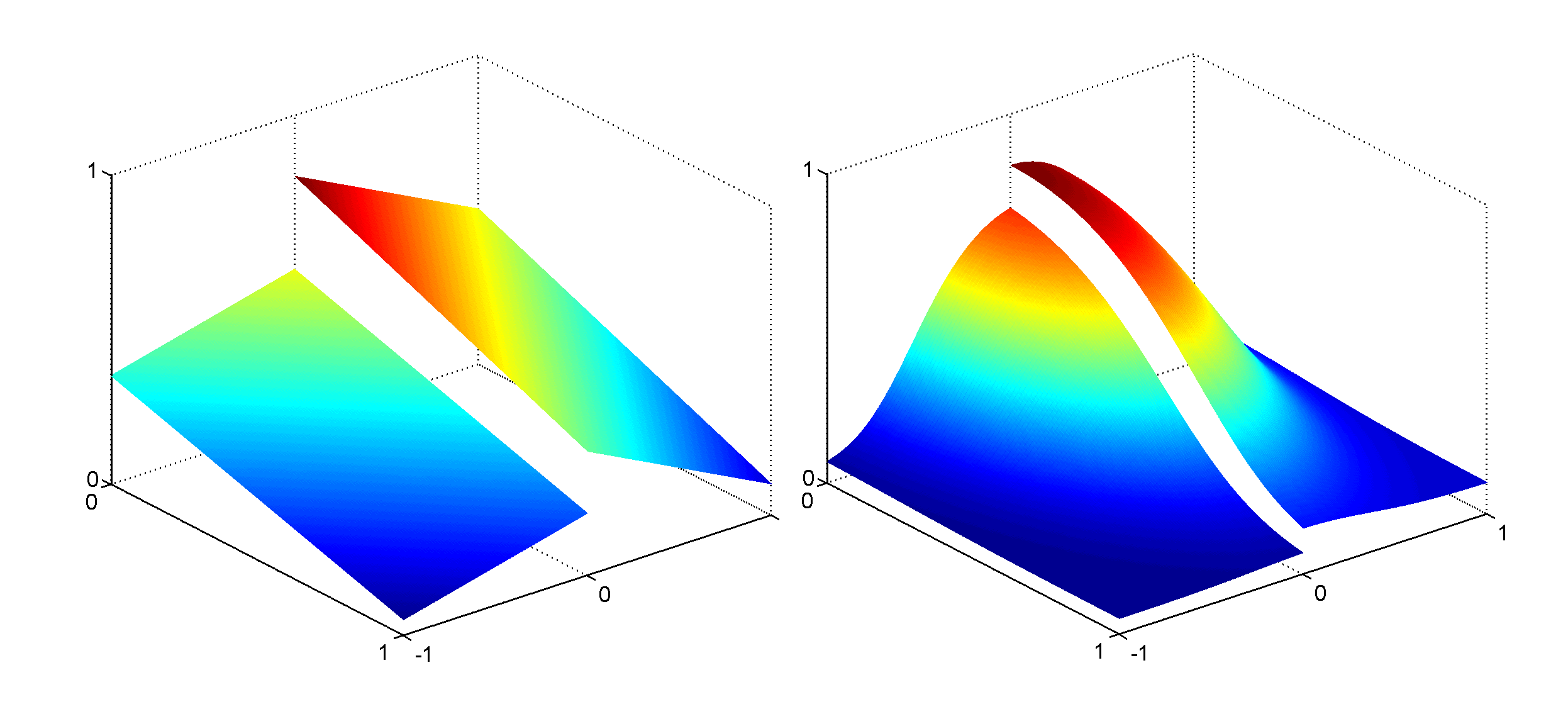}
    \caption{A illustration of possible reconstructions by the MUSCL scheme (left) and the THINC/QQ scheme (right)}
    \label{fig:muscl-thincqq}
\end{figure}

\subsubsection{The one-stage BVD scheme}

Similar to the MUSCL-THINC-BVD scheme for structured grids, the one-stage BVD scheme has two candidate schemes: a MUSCL scheme and a THINC/QQ scheme with a relatively large steepness parameter $\beta_l$. The MUSCL scheme
 tends to smear discontinuities over several grids due to numerical dissipation. On the other hand, The THINC/QQ scheme can preserve the solution structure of discontinuities even for long-term simulations. Thus, using the BVD algorithm as the selecting procedure, we implement the MUSCL scheme at smooth regions and the THINC/QQ scheme at discontinuities.
According to \cite{deng2018high,deng2018limiter} and our numerical tests, $\beta_l \sim (1.3,1.6)$ can give good or acceptable results and $\beta_l=1.4$ is used for all following test cases.

The one-stage BVD scheme is formulated as
\begin{equation}
    q_{ij,\Theta} = \left\{  \begin{array}{ll}
                        q_{ij,\Theta}^l  & \textrm{if } TBV_i^l < TBV_i^m \\
                        q_{ij,\Theta}^m  &  \textrm{otherwise}
                             \end{array}
                    \right. ,
    \Theta = \left\{ \begin{array}{ll}
                  L  & \textrm{if } \bm{n}_{ij} \textrm{ points outside of } \Omega_i \\
                  R  & \textrm{otherwise} 
                     \end{array}
             \right.,
    j=1,2,\dots,J
\end{equation}
where $TBV_i^{m/l}$ are $TBV$s corresponding to the MUSCL scheme and the THINC/QQ scheme with $\beta_l$.

Numerical tests show that this one-stage BVD scheme can efficiently reduce numerical diffusion at strong discontinuities such as shock waves and material interfaces. However, since a relatively large steepness $\beta_l$ is used, it can not capture some slight discontinuities such as shear waves and vortices.

\subsubsection{The two-stage BVD scheme}

The two-stage BVD scheme was inspired by a further investigation into the THINC/QQ scheme. Figure \ref{fig:muscl-thincqq} shows possible reconstructions in two quadrilateral cells by the MUSCL scheme and the THINC/QQ scheme. As mentioned previously, The THINC/QQ function of Eq. \ref{eq:thincqq} includes not only gradient terms but also curvature terms. Thus, it can preserve properties of some curved flow structures such as vortices better than the MUSCL scheme. To improve the performance of the one-stage BVD scheme at those curved smooth regions, and also for slight discontinuities, we add another THINC/QQ reconstruction with a relatively small steepness $\beta_s$ as the third candidate. According to \cite{deng2018limiter} and our numerical tests, $\beta_s \sim (0.7,1.0)$ can give good or acceptable results. A $\beta_s=0.8$ is used for all the following test cases.

In the two-stage BVD scheme, we have three group of candidate states, say, $q_{ij,L/R}^{I}(I=m,s,l)$ from the MUSCL scheme, the THINC/QQ scheme with $\beta_s$ and the THINC/QQ scheme with $\beta_l$. The final reconstruction are chosen according to the following criterion.
\begin{equation}
    q_{ij,\Theta} = \left\{  \begin{array}{ll}
                        q_{ij,\Theta}^l  & \textrm{if } TBV_i^l < TBV_i^m \textrm{and } TBV_i^l < TBV_i^s \\
                        q_{ij,\Theta}^s  & \textrm{if } TBV_i^s < TBV_i^m \textrm{and } TBV_i^s < TBV_i^l \\
                        q_{ij,\Theta}^m  &  \textrm{otherwise}
                             \end{array}
                    \right. ,
    \Theta = \left\{ \begin{array}{ll}
                  L  & \textrm{if } \bm{n}_{ij} \textrm{ points outside of } \Omega_i \\
                  R  & \textrm{otherwise} 
                     \end{array}
            \right.,
    j=1,2,\dots,J
\end{equation}

It is noticed that, this two-stage BVD algorithm is simpler than the one for structured grids in \cite{deng2019fifth}. In fact, we compare all three $TBV$s directly and choose the scheme with the smallest $TBV$. This means that our two stages are parallel with a uniform algorithm. But for the structured one, two stages have a determined order to treat smooth regions and discontinuities separately.

\section{Numerical results}

\subsection{Solid rotation of a complex profile}
\begin{figure}[htbp]
    \begin{minipage}[t]{0.49\linewidth}
        \centering
        \includegraphics[width=0.9\textwidth]{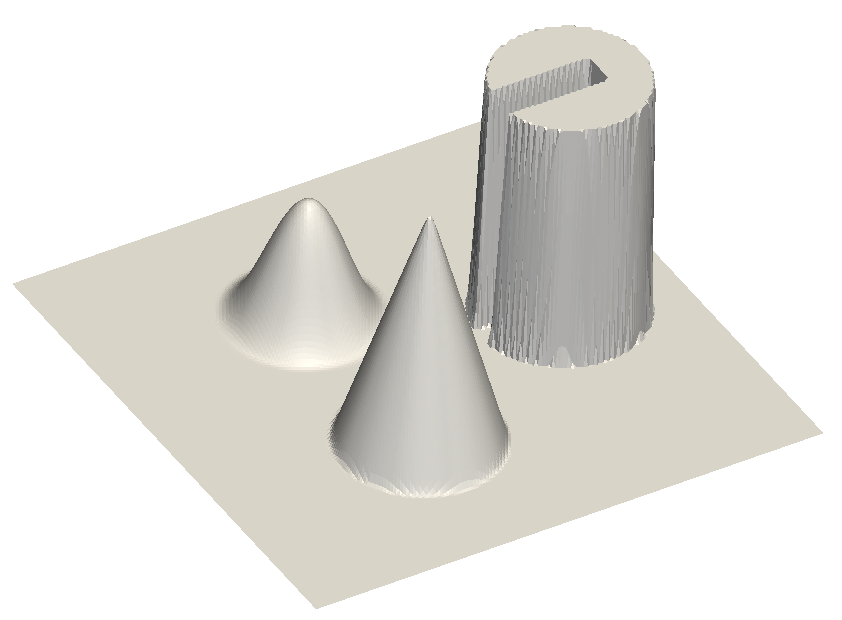}
        \caption{The initial condition of the rotation test}
        \label{fig:t1:init}
    \end{minipage}
    \begin{minipage}[t]{0.49\linewidth}
        \centering
        \includegraphics[width=0.9\textwidth]{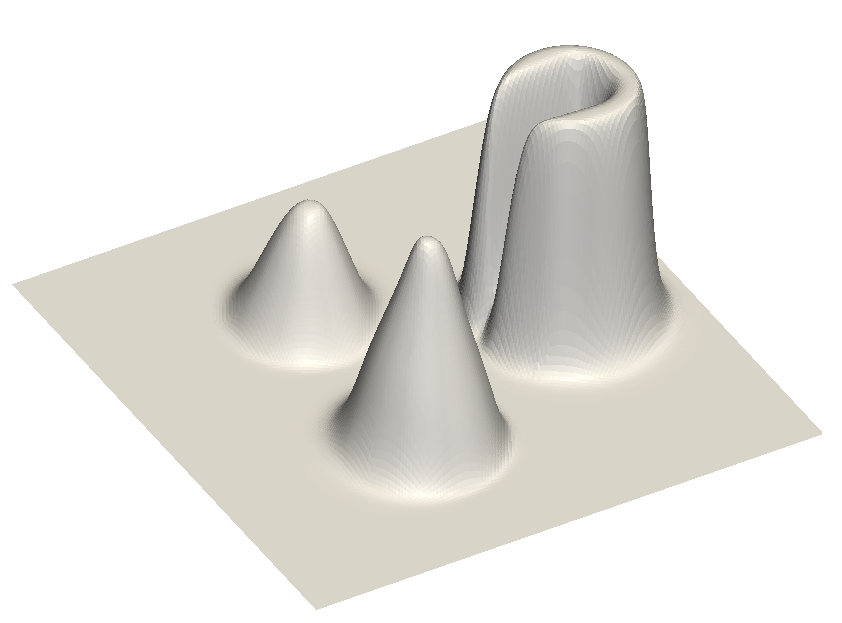}
        \caption{The solution of the MUSCL scheme}
        \label{fig:t1:muscl}
    \end{minipage}
\end{figure}

This case can assess the ability of present schemes to resolve sharp discontinuities and keep smooth regions smooth. We consider a complex 2D benchmark problem used by \cite{leveque1996high,xie2019high} .The computational domain is $[0,1]^{2}$, divided into $54,604$  triangular elements. The initial condition consists three shapes within three circles of radius $r_0=0.15$ respectively, as showed in Figure \ref{fig:t1:init}.
\begin{displaymath}
\phi = \left\{ \begin{array}{ll}
    1, \textrm{ if} \left|x-0.5\right|<0.25, \textrm{ or }y>0.85 &  r_1(x,y)<r_0\\
    \frac{1}{4}(1+\textrm{cos}(\pi\textrm{min}(r_2(x,y)/r_0,1))) & r_2(x,y)<r_0\\
    1-r_3(x,y)/r_0 & r_3(x,y)<r_0 \\
    0 & \textrm{otherwise}
    \end{array}\right.
\end{displaymath}
where $r_i(x,y)=\sqrt{(x-x_i)^{2}+(y-y_i)^{2}}$, $(x_1,y_1)=(0.5,0.78)$, $(x_2,y_2)=(0.31,0.39)$, $(x_3,y_3)=(0.69,0.39)$.

\begin{figure}[htbp]
    \centering
    \subfigure[]
    {
        \centering
        \includegraphics[width=0.45\textwidth]{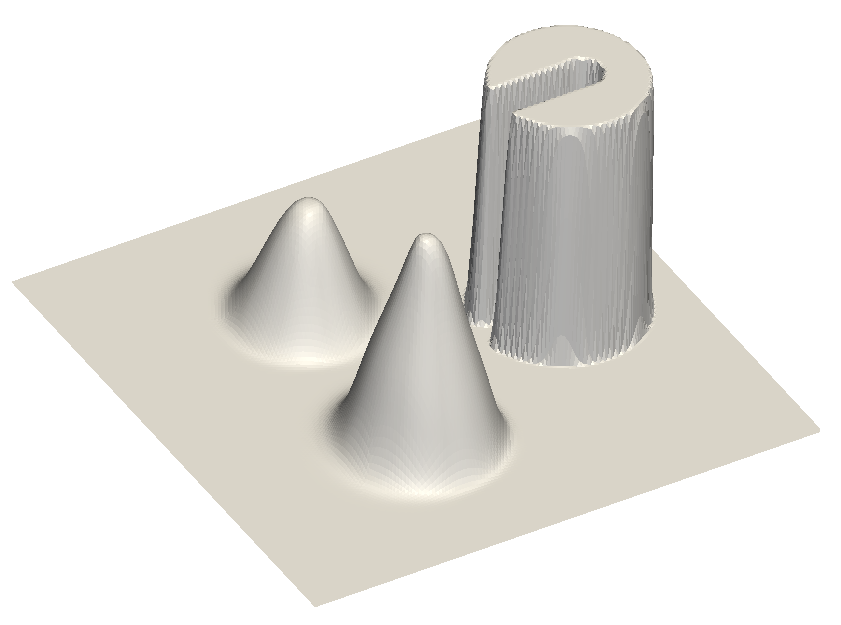}
        \hspace{0.5cm}
    }
    \subfigure[]
    {
        \centering
        \includegraphics[width=0.45\textwidth]{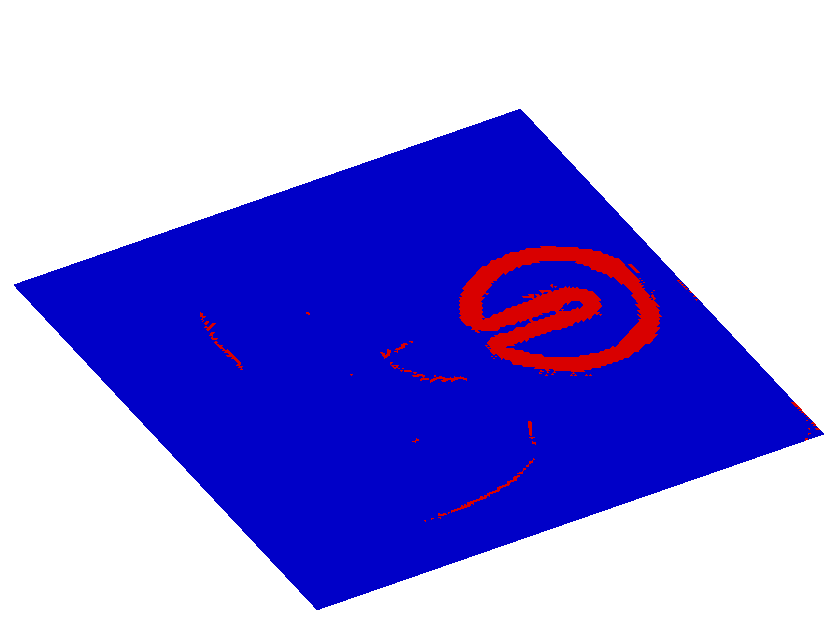}
    }
    \caption{Results of the one-stage BVD scheme: (a)The solution after one rotation; (b)Red cells use THINC/QQ scheme}
    \label{fig:t1:bvd}
\end{figure}

\begin{figure}[htbp]
    \centering
    \subfigure[]
    {
        \centering
        \includegraphics[width=0.45\textwidth]{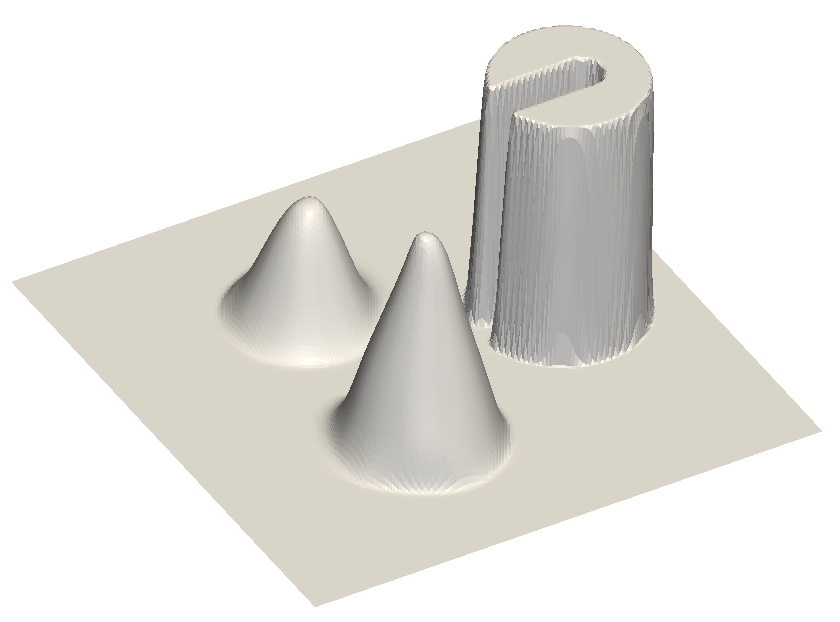}
        \hspace{0.5cm}
    }
    \subfigure[]
    {
        \centering
        \includegraphics[width=0.45\textwidth]{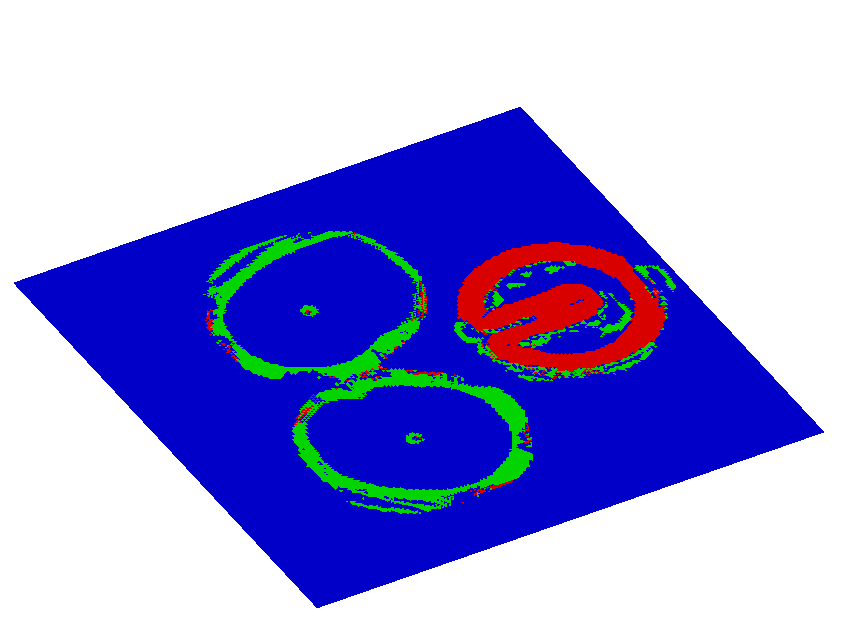}
    }
    \caption{Results of the two-stage BVD scheme: (a)The solution after one rotation; (b)Red and yellow cells use THINC/QQ scheme}
    \label{fig:t1:tsbvd}
\end{figure}

It includes a classical shape of a slotted disk proposed by Zalesak \cite{zalesak1979fully}, a smooth hump and a sharp cone. These shapes are rotated by a velocity field of 
\begin{displaymath}
u = 0.5-y, v = x-0.5
\end{displaymath}
Zero-gradient boundary condition is applied to all boundaries. Results of $t=2\pi$ by different schemes are showed in Figure \ref{fig:t1:muscl} to \ref{fig:t1:tsbvd}. The maximum Courant number is set to $0.2$.

For the smooth hump and cone, three schemes give almost the same results. Since BVD schemes choose the MUSCL scheme for these region, as showed in Figure \ref{fig:t1:bvd}(b) and Figure \ref{fig:t1:tsbvd}(b). Although the two-stage BVD scheme chooses THINC/QQ scheme with $\beta_s$ in some cells (marked with yellow color) around hump and cone, it almost gives the same solution as the MUSCL scheme. On the other hand, BVD schemes can find cells including discontinuity (marked with red color), like the edge of the slotted disk, and implement THINC/QQ scheme with $\beta_l$ which can preserve the step-like solution structure. This step-like solution is smeared by the pure MUSCL scheme, as showed in Figure \ref{fig:t1:muscl}.

\subsection{Gas Dynamic Problems}
In this section, we assess the ability of BVD schemes to capture shock waves and vortices in gas dynamic problems. The Euler equation is solved by prescribed methods with the HLL Riemann solver. The ratio of specific heats is $\gamma = 1.4$.

\subsubsection{A Riemann Problem}
Two-dimensional schemes are applied to a one-dimensional shock tube problem. The computational domain is $[0,1]\times[0,0.1]$ with $100$ triangular elements in the $x$-direction and $10$ triangular elements in the $y$-direction. We consider the following Riemann-type initial conditions \cite{toro2013riemann}:
\begin{displaymath}
(\rho,u,p)=\left\{ \begin{array}{ll}
    (1.0, 0.0, 1000.0) & \textrm{if } x<0.5 \\
    (1.0, 0.0, 0.01) & \textrm{otherwise}
\end{array}
\right.
\end{displaymath}

\begin{figure}[htbp]
    \centering
    \subfigure[]
    {
        \centering
        \includegraphics[width=0.47\textwidth]{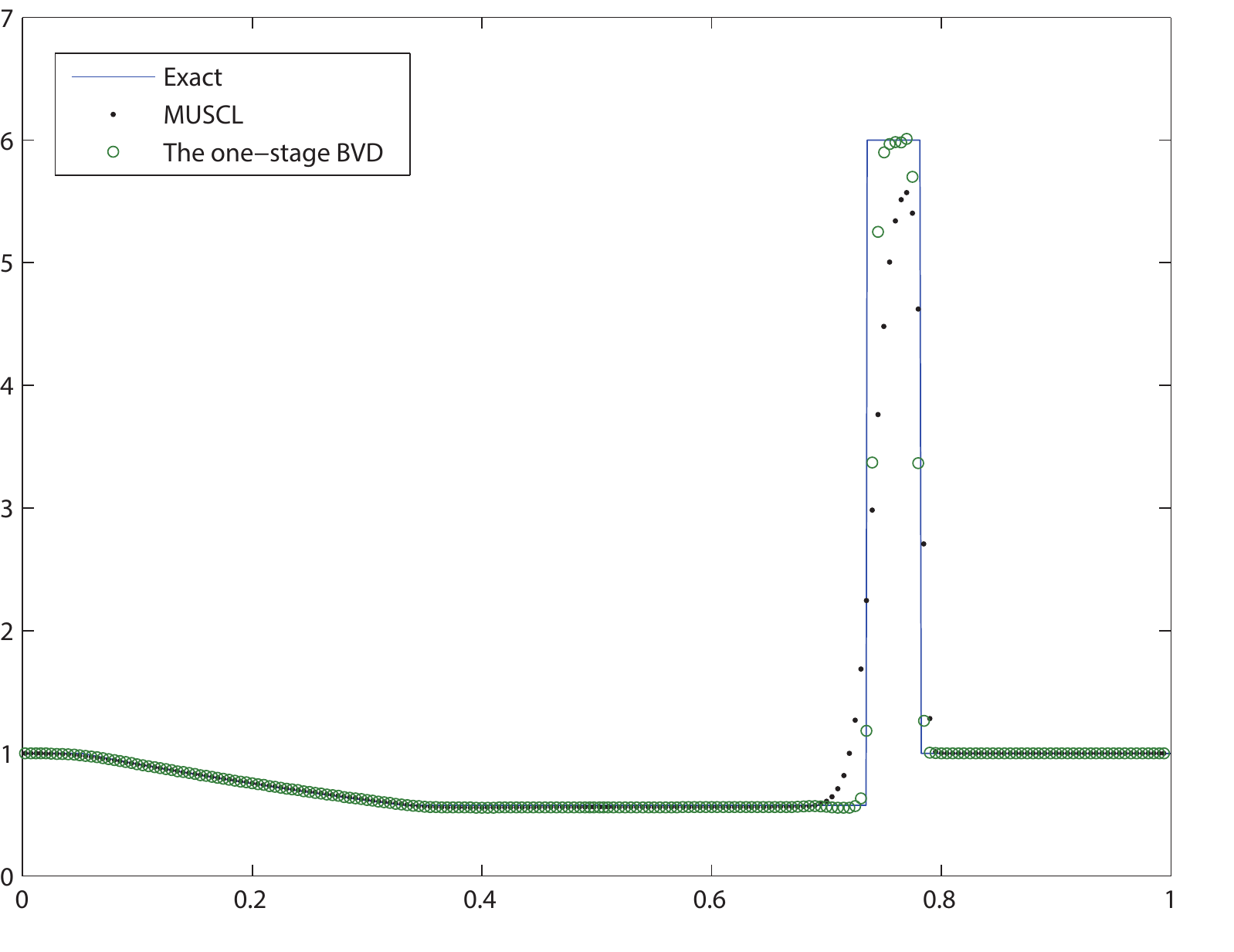}
        \hspace{0.3cm}
    }
    \subfigure[]
    {
        \centering
        \includegraphics[width=0.47\textwidth]{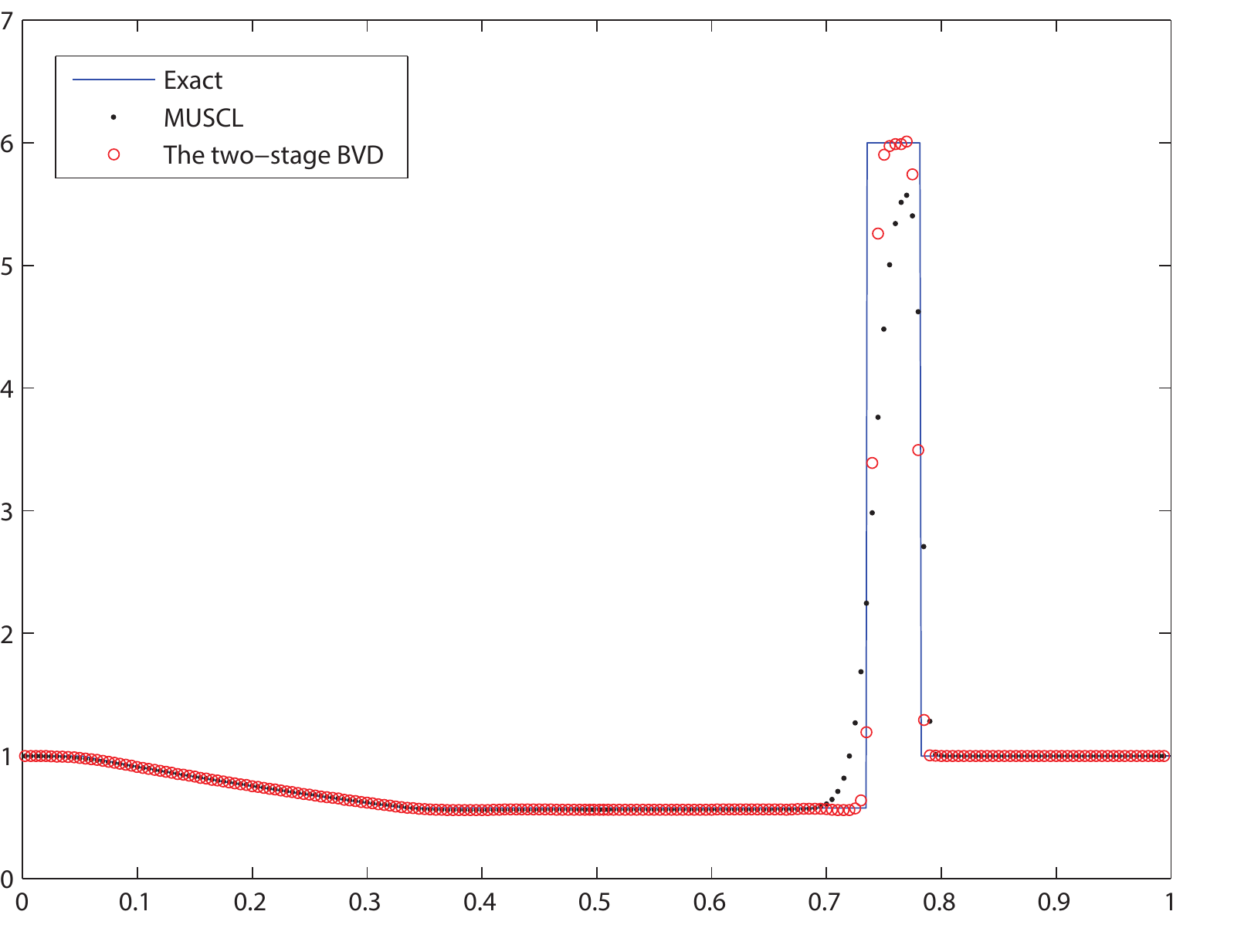}
    }
    \caption{Density of the Riemann problem at $t=0.012$}
    \label{fig:t2:density}
\end{figure}

\begin{figure}[htbp]
    \centering
    \subfigure[The one-stage BVD scheme]
    {
        \centering
        \includegraphics[width=0.8\textwidth]{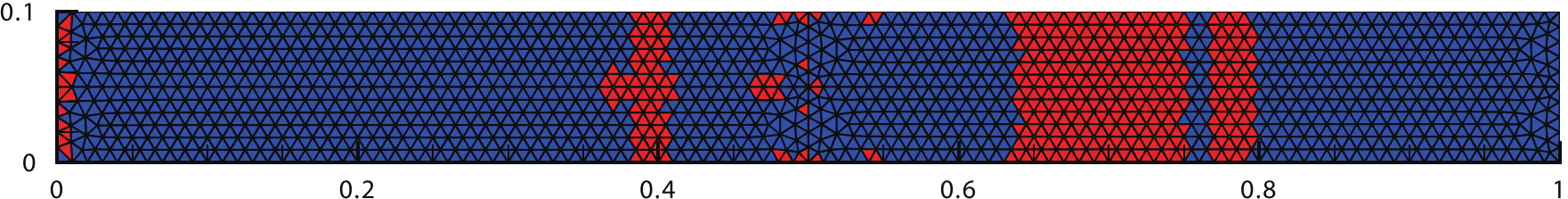}
    }
    \subfigure[The two-stage BVD scheme]
    {
        \centering
        \includegraphics[width=0.8\textwidth]{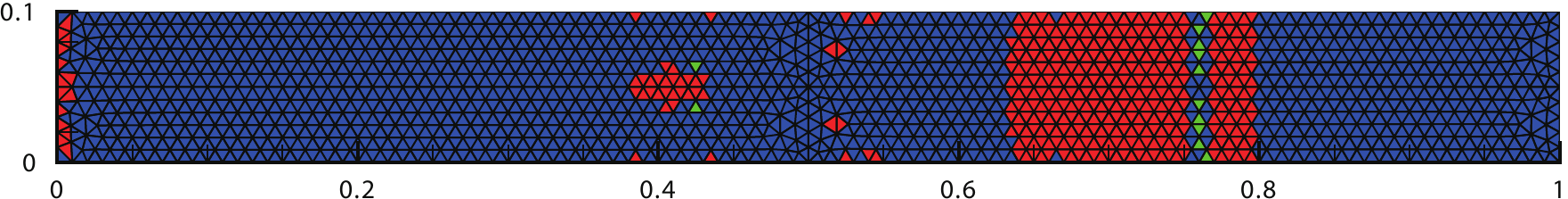}
    }
    \caption{Cells using THINC/QQ schemes to reconstruct density}
    \label{fig:t2:wcqr}
\end{figure}

It is the left half of the blast wave problem of Woodward and Colella \cite{woodward1984numerical}. Its solution contains a left rarefaction, a contact wave and a right-moving shock wave. Density of the exact solution and numerical methods at $t=0.012$ are showed in Figure \ref{fig:t2:density}. It is obvious that BVD schemes can resolve a sharper shock and contact wave than the pure MUSCL scheme, and there is no oscillation around discontinuities. Figure \ref{fig:t2:wcqr} shows cells using THINC/QQ schemes for reconstructing $\rho$ when BVD schemes are implemented. For cells around the shock and contact waves, the THINC/QQ with $\beta_l$ is used (red cells), which can preserve the step-like flow structure.

\subsubsection{A Mach 3 Wind Tunnel with a Step}
\begin{figure}
    \centering
    \includegraphics[width=0.7\textwidth]{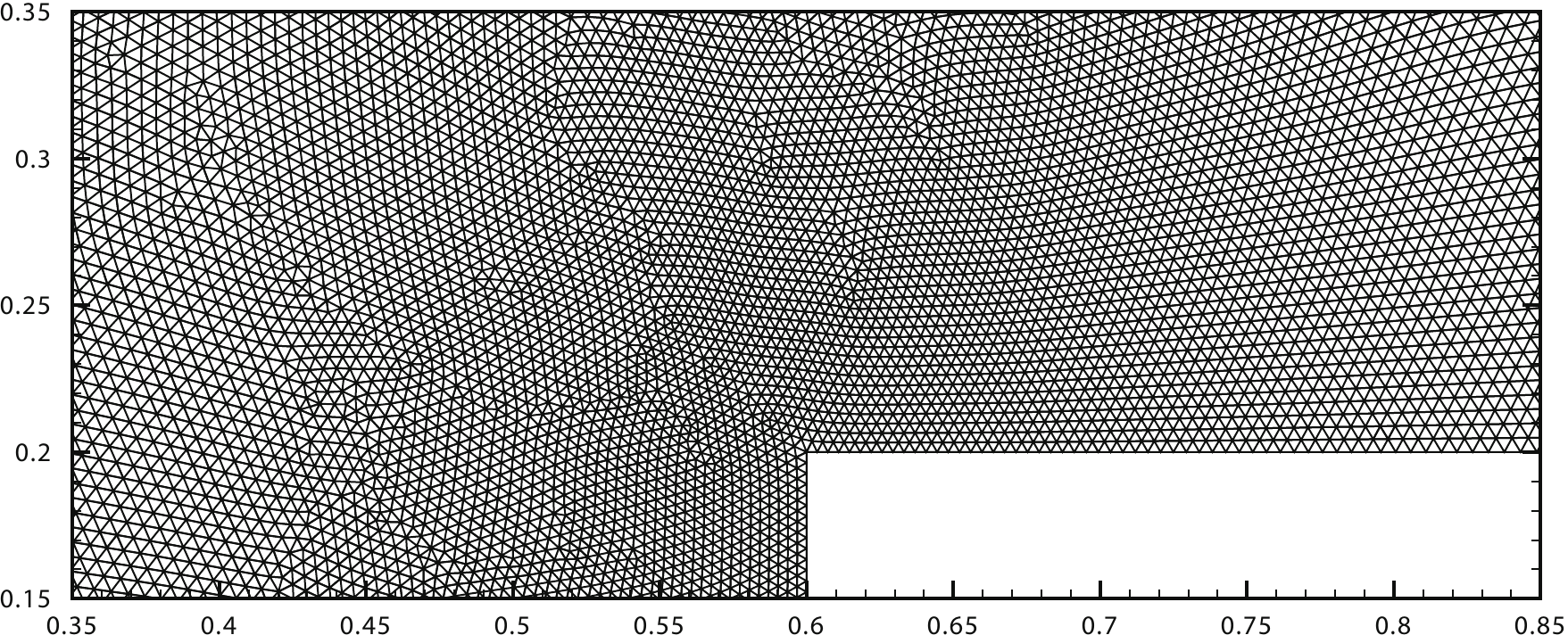}
    \caption{Mesh of wind tunnel around the corner}
    \label{fig:t3:mesh}
\end{figure}

This problem is also from \cite{woodward1984numerical}, and widely used to verify the capability of numerical schemes in capturing strong shocks and vortices \cite{hu1999weighted,li2012high,wolf2007high,zhu2008runge,zhu2009hermite}. The wind tunnel is $1.0$ unit high and $3.0$ unit long. A step is located $0.6$ unit from the left boundary with a height of $0.2$ unit. This domain is divided into triangular elements with a size of $1/160$ away from the corner but $1/320$ around the corner, as Figure \ref{fig:t3:mesh}. This mesh was used by \cite{hu1999weighted} to deal with the singularity point of the corner. The initial condition is the same as the inflow passing through the scram-jet engine.

\begin{figure}[htbp]
    \centering
    \subfigure[The muscl scheme]
    {
        \centering
        \includegraphics[width=0.48\textwidth]{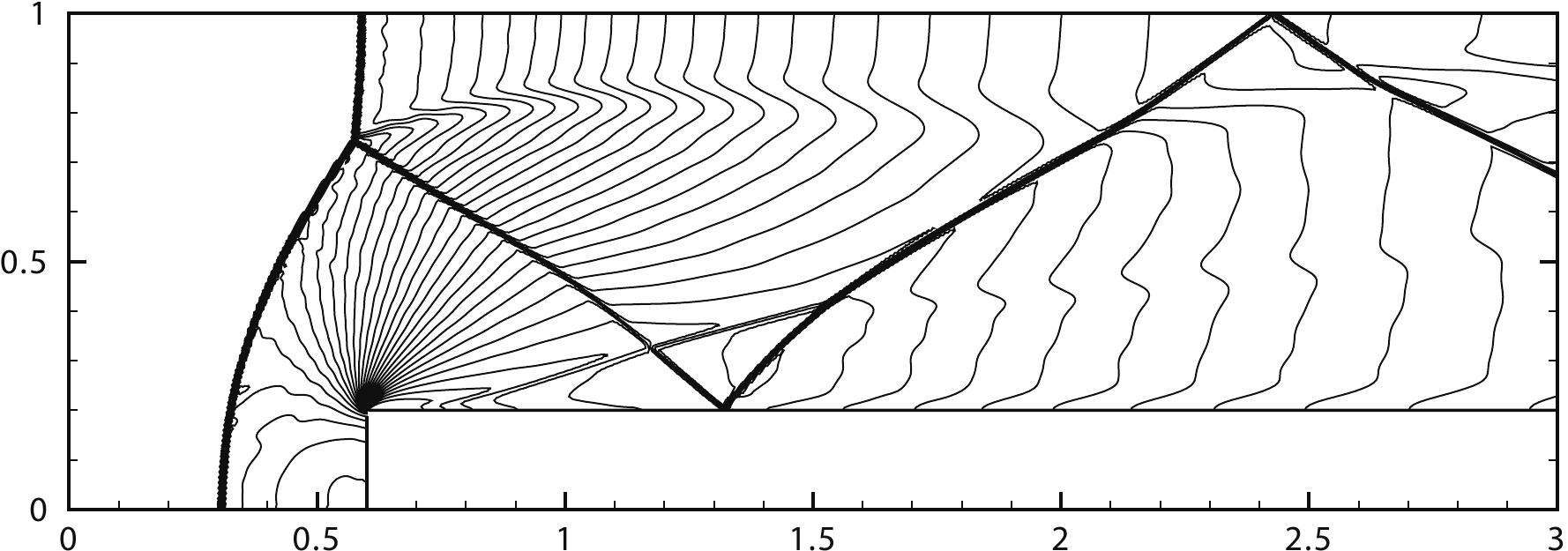}
    }
    \subfigure[The one-stage BVD scheme]
    {
        \centering
        \includegraphics[width=0.48\textwidth]{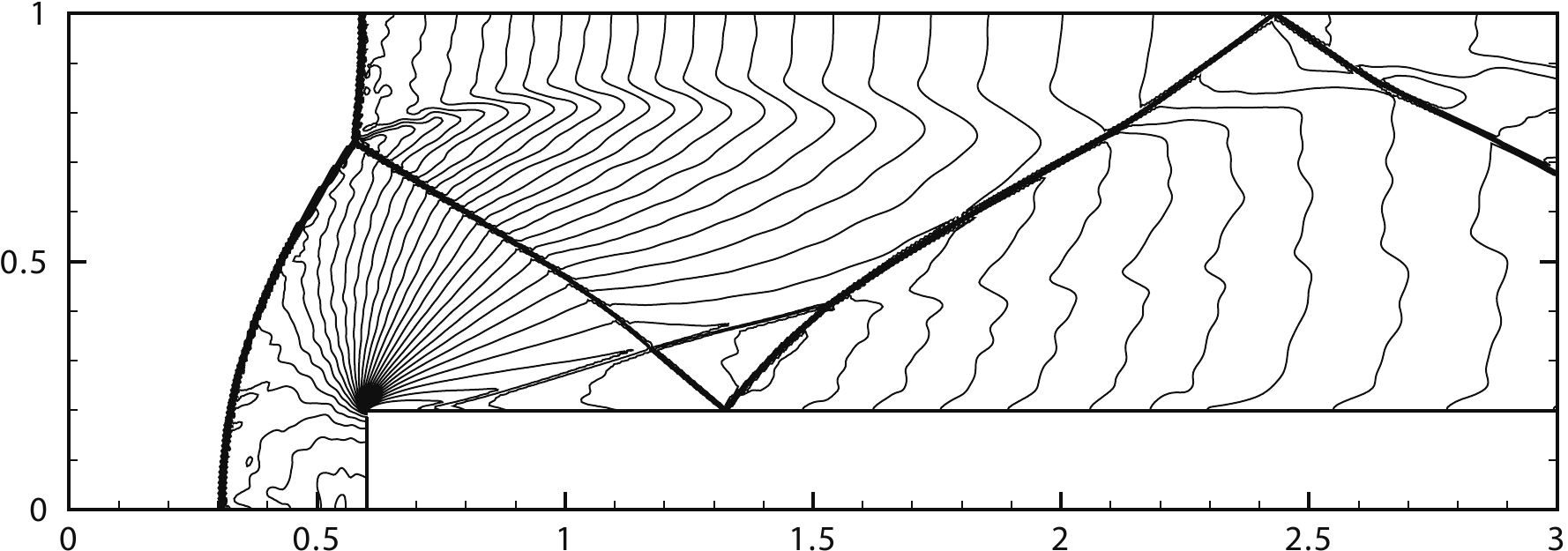}
    }
    \subfigure[The two-stage BVD scheme]
    {
        \centering
        \includegraphics[width=0.48\textwidth]{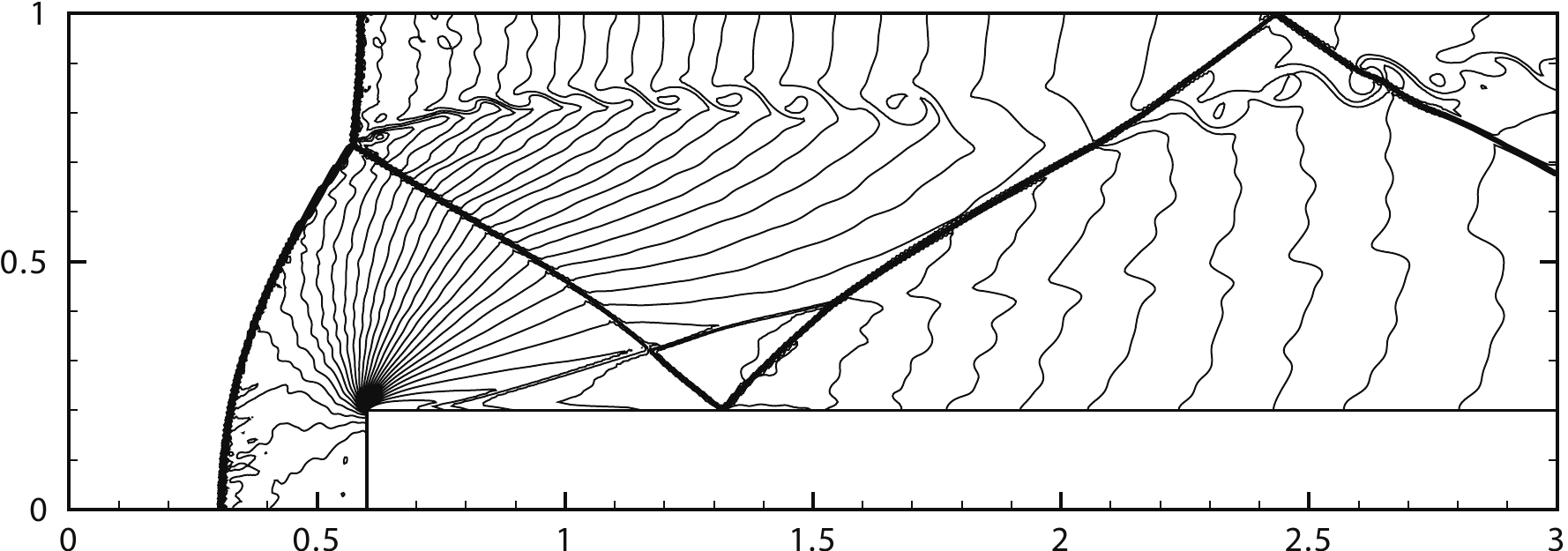}
    }
    \subfigure[The result of a 3rd-order scheme \cite{hu1999weighted} with the same grid]
    {
        \centering
        \includegraphics[width=0.48\textwidth]{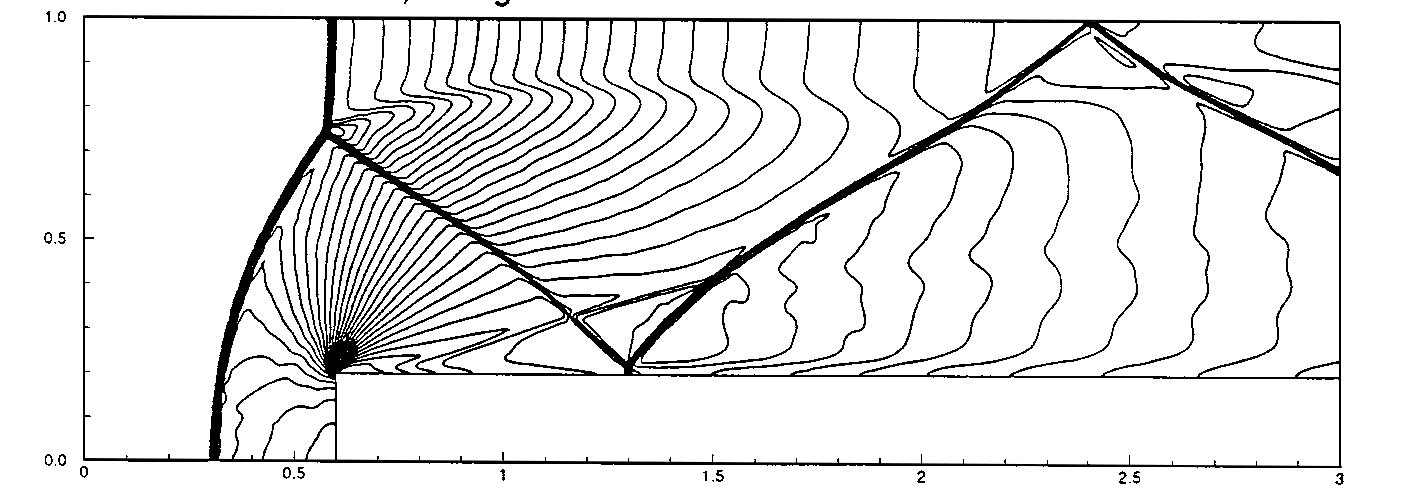}
    }
    \caption{Density contour of results at $t=4.0$}
    \label{fig:t3:cont}
\end{figure}

\begin{figure}[htbp]
    \centering
    \subfigure[The one-stage BVD scheme]
    {
        \centering
        \includegraphics[width=0.48\textwidth]{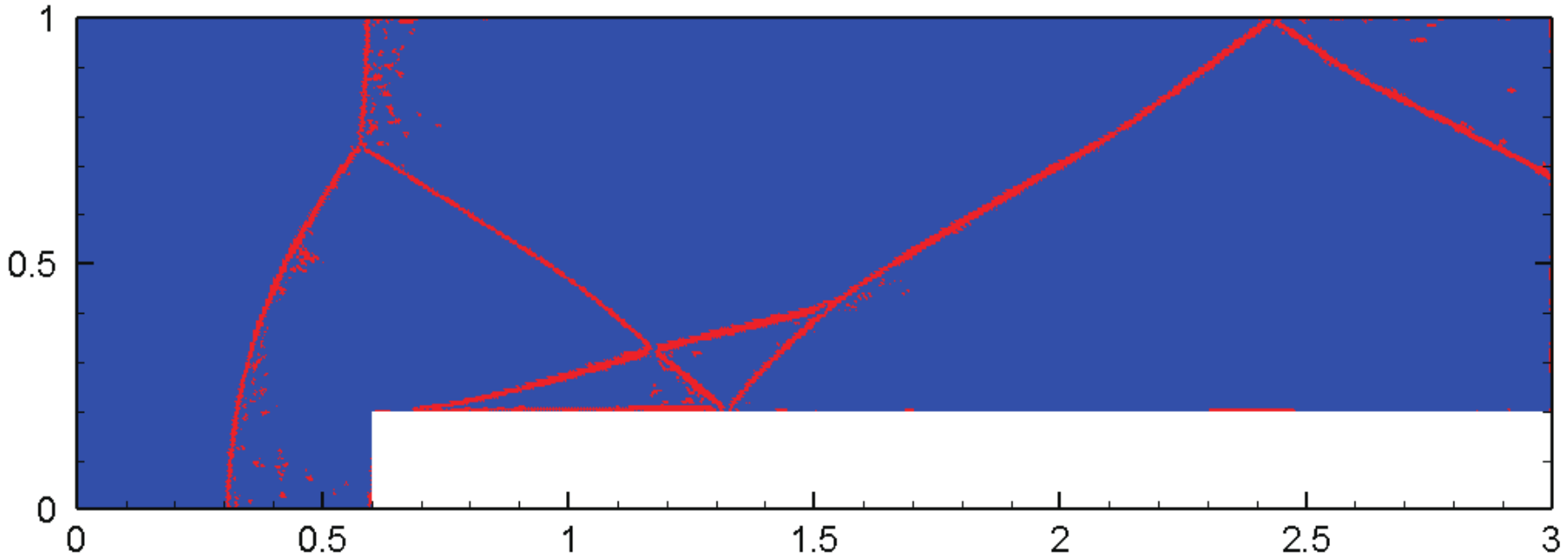}
    }
    \subfigure[The two-stage BVD scheme]
    {
        \centering
        \includegraphics[width=0.48\textwidth]{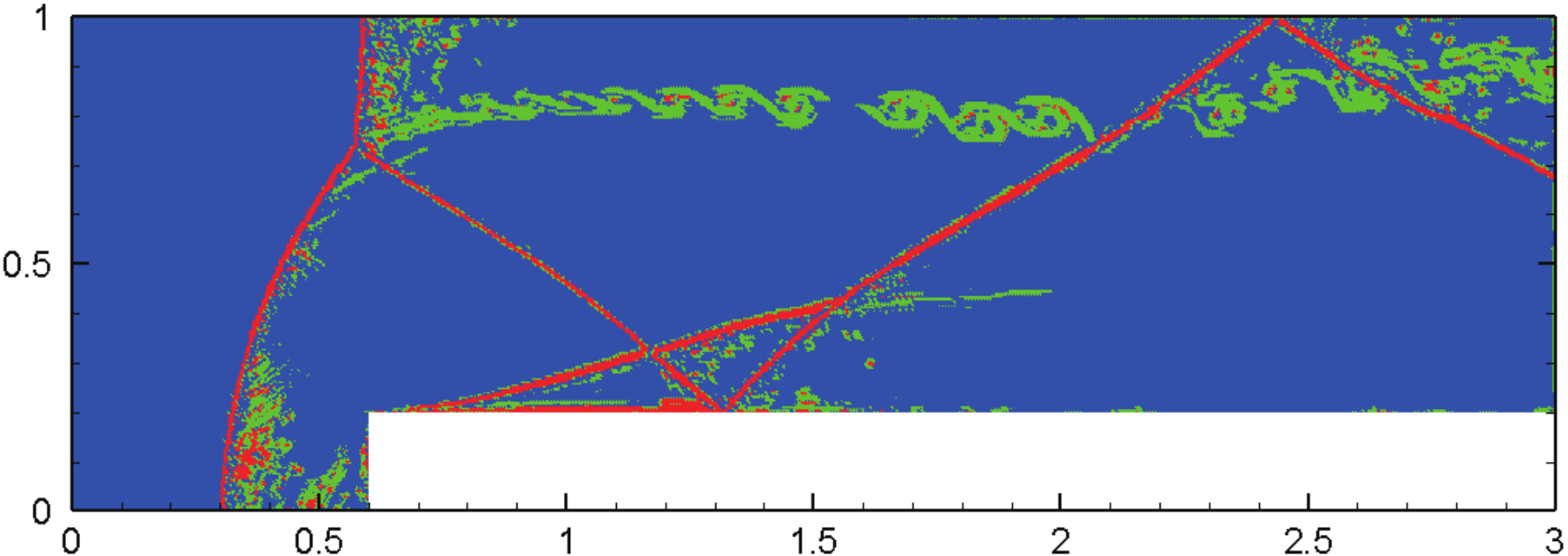}
    }
    \caption{Cells using THINC/QQ schemes to reconstruct $\rho$}
    \label{fig:t3:wr}
\end{figure}

\begin{figure}[htbp]
    \centering
    \subfigure[The one-stage BVD scheme]
    {
        \centering
        \includegraphics[width=0.48\textwidth]{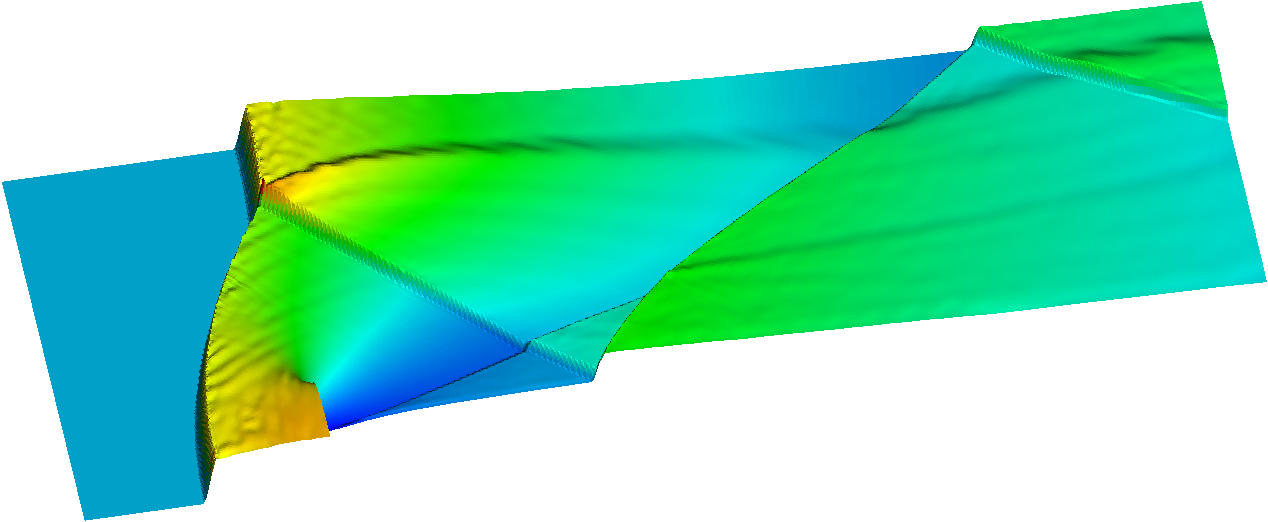}
    }
    \subfigure[The two-stage BVD scheme]
    {
        \centering
        \includegraphics[width=0.48\textwidth]{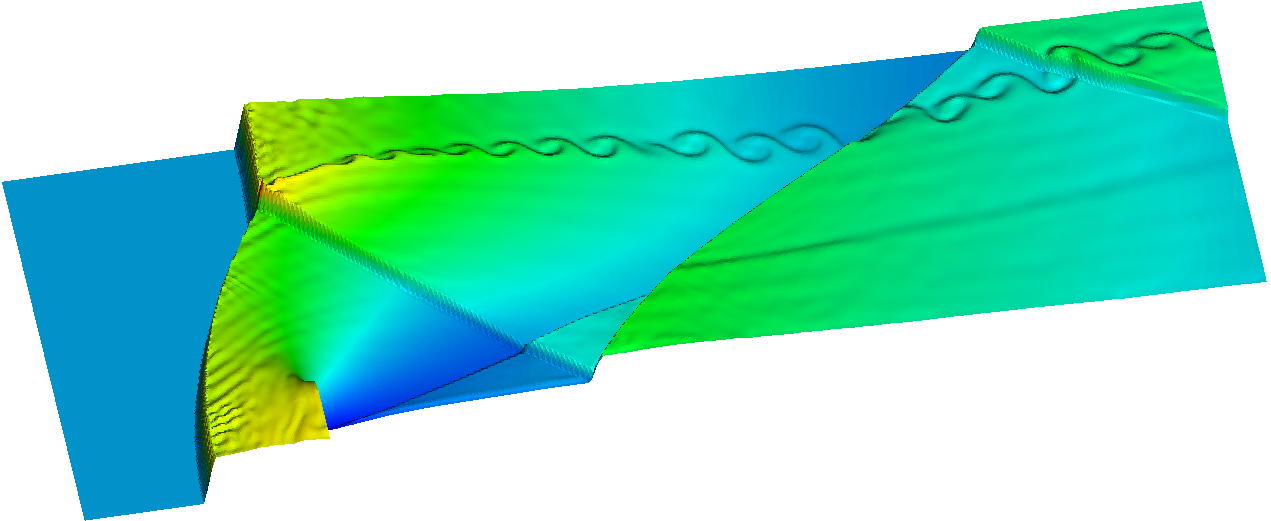}
    }
    \caption{Three-dimensional bird's view of the density field}
    \label{fig:t3:3d}
\end{figure}

Figure \ref{fig:t3:cont} is the density contour at $t=4.0$, ranging from 0.32 to 6.15 with 30 equivalent intervals. It is seen that with the help of THINC/QQ schemes, BVD schemes can resolve shock waves better than pure MUSCL scheme. Red cells in FIgure \ref{fig:t3:wr} use THINC/QQ with $\beta_l$. They are also cells including shock waves. On the other hand, the two-stage BVD scheme can resolve very clear vortices, as showed in Figure \ref{fig:t3:3d}. This benefits from the implementation of THINC/QQ with $\beta_s$, which also includes curvature information in second order terms but not as steep as the one with $\beta_l$. It can preserve flow structures of vortices. By comparing with the result of a 3rd-order WENO scheme \cite{hu1999weighted}, we can see that although the order of both the MUSCL scheme and the THINC/QQ scheme \cite{xie2017toward} are around 2nd-order, the combination of them by BVD algorithms can sometimes give better results than a 3rd-order scheme.


\subsection{Two-fluid flow problems}
One of main applications of the THINC scheme and the THINC/QQ scheme is to capture immersed interfaces. They can limit interfaces into several cells even over very long-term simulations \cite{xiao2005simple,xie2017toward,shyue2014eulerian,pandare2019enhanced}. The combination of the THINC sheme and the MUSCL scheme by BVD algorithm in the structured-gird framework \cite{deng2018high} showed that it can not only resolve a very clear interface, but also capture more flow details than traditional polynomial reconstructions such as the MUSCL scheme and the WENO sheme. In this section, we will see that the MUSCL-THINC/QQ-BVD scheme in the unstructured-grid framework has comparable performance. The five-equation model is solved by prescribed numerical schemes with the HLLC Riemann solver.

\subsubsection{Two dimensional shock--R22-cylinder interaction}
\begin{figure}[htbp]
    \centering
    \includegraphics[width=0.7\textwidth]{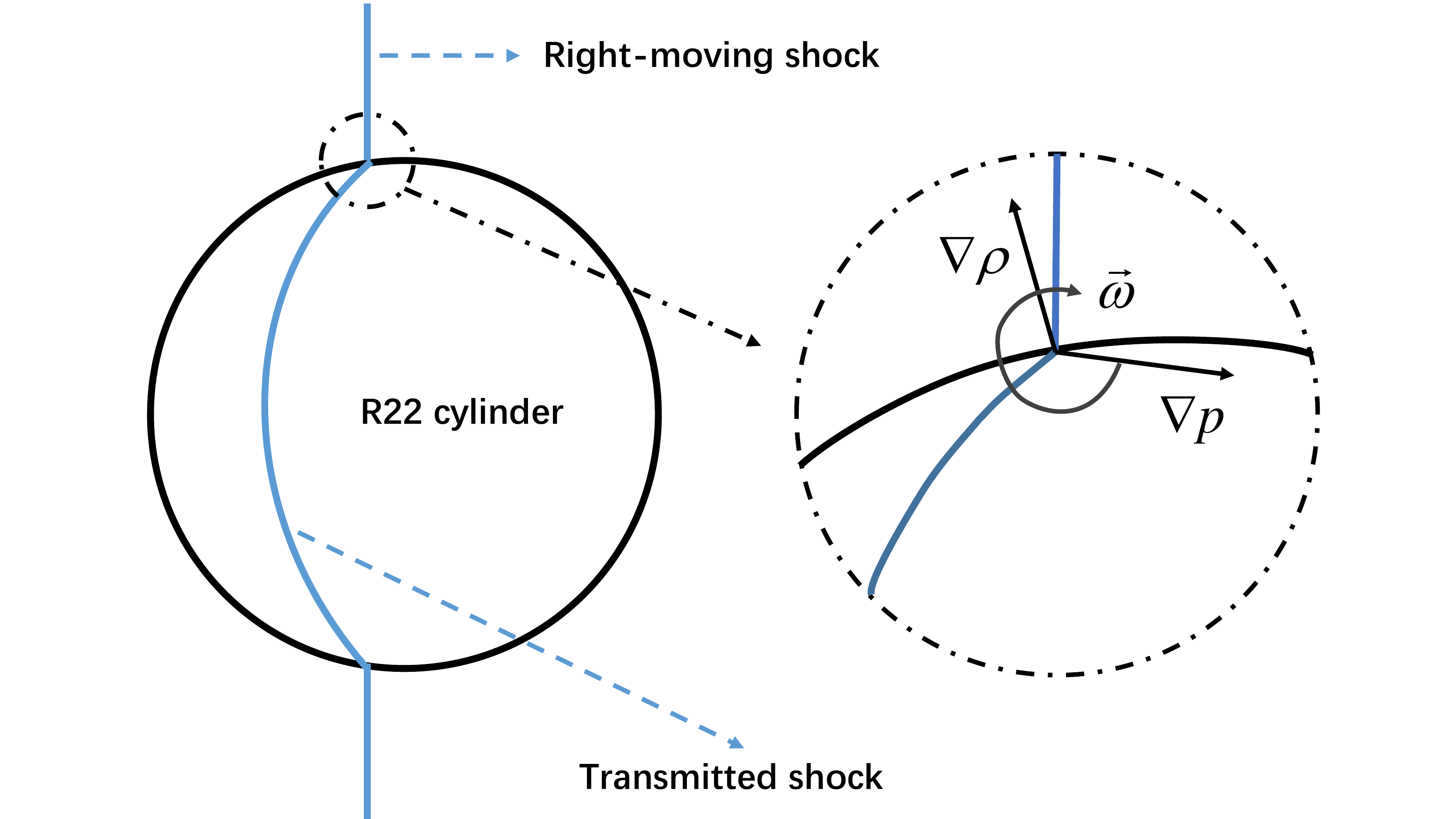}
    \caption{Generation of vorticity in interaction between air shock and R22 cylinder}
    \label{fig:t6:vortex}
\end{figure}

As the first benchmark test of two-component flows, we consider a well known shock-bubble interaction problem involving interaction between a shock in air and a R22 cylinder \cite{deng2018high,quirk1996dynamics,shyue2006wave,shankar2010numerical,so2012anti}. As analyzed in \cite{picone1988vorticity,giordano2006richtmyer}, vortices will be generated by the baroclinic mechanism when the shock wave pass through the surface of R22 cylinder. Consider the equation of vorticity without viscous terms proposed in \cite{picone1988vorticity}:
\begin{equation} \label{eq:vortex}
    \frac{d\bm{\omega}}{dt}+\bm{\omega} \nabla \cdot \bm{V} = \bm{\omega} \cdot \nabla \bm{V} + \frac{\nabla\rho\times\nabla p}{\rho^2}
\end{equation}
The last term of Eq.(\ref{eq:vortex}) is a source term. Misalignment of the local gradient of pressure and local gradient of density will lead to a generation of vorticity. Figure \ref{fig:t6:vortex} shows the direction of verticity generated in the interaction between a right-moving air shock and a R22 cylinder. If the numerical diffusion of a numerical scheme is too strong, it will smear these vortices.
\begin{figure}[htbp]
    \centering
    \includegraphics[width=0.7\textwidth]{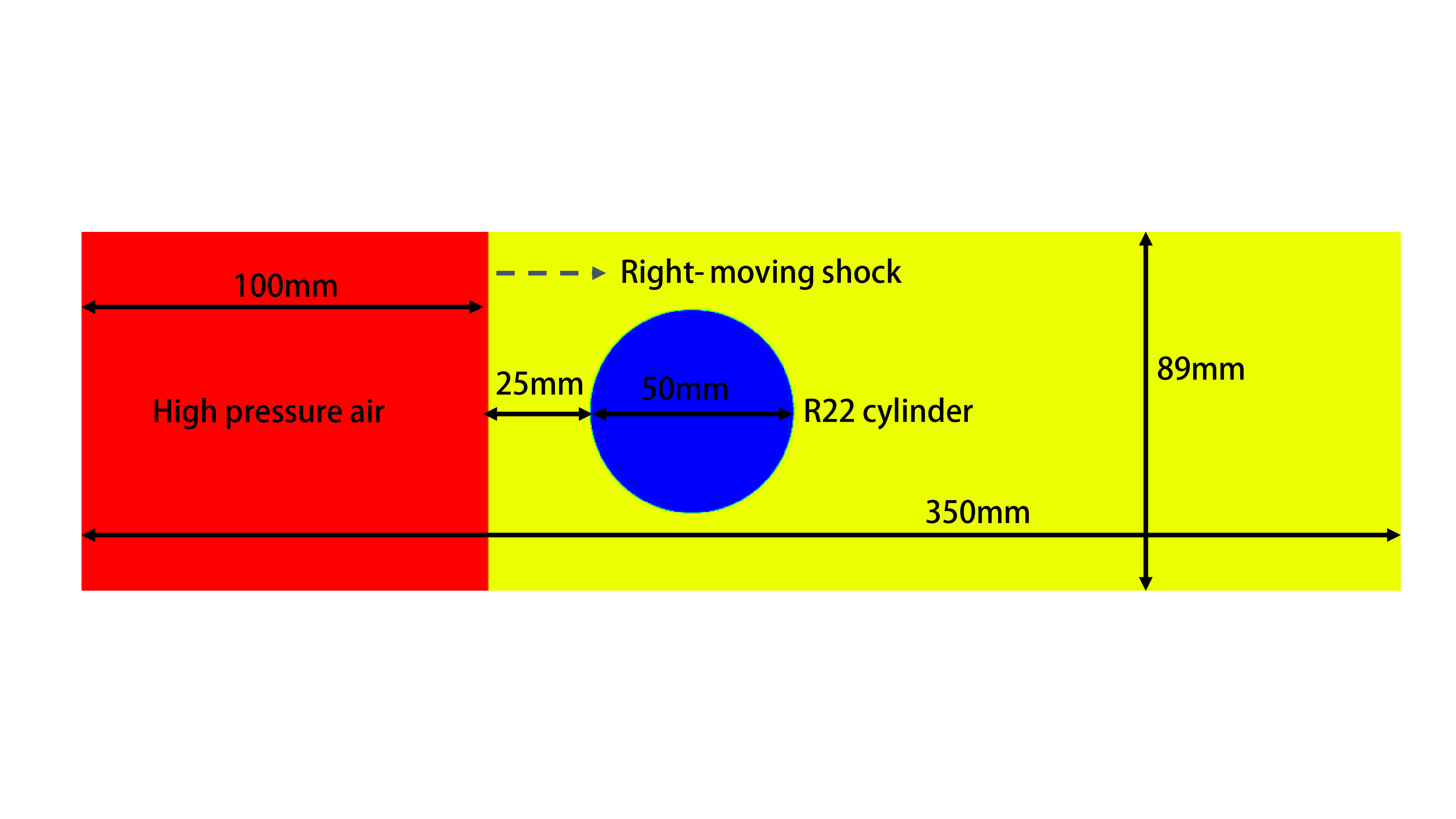}
    \caption{The computational domain of shock--R22-cylinder interaction}
    \label{fig:t6:init}
\end{figure}

The computational setup is shown in Figure \ref{fig:t6:init}. Refer to \cite{haas1987interaction} for experimental results. A planar right-moving Mach 1.22 shock in air hits a stationary R22 gas cylinder with a diameter $d=50mm$. Both air and R22 gas are treated as ideal gases. The initial condition is given as:
\begin{displaymath}
(\alpha_1,\rho,u,v,p,\gamma)=\left\{ \begin{array}{ll}
    (\epsilon,3.863 kg/m^3, 0.0, 0.0, 1.01325 \times 10^5 Pa, 1.249) & \textrm{In the R22 cylinder} \\
    (1.0-\epsilon,1.686 kg/m^3, 113.5 m/s, 0.0, 1.59 \times 10^5 Pa, 1.4) & \textrm{Post-shock} \\
    (1.0-\epsilon,1.225 kg/m^3, 0.0, 0.0, 1.01325 \times 10^5 Pa, 1.4) & \textrm{Otherwise}
\end{array}
\right.
\end{displaymath}
where $\epsilon=10^{-8}$. A uniform triangular mesh with $h=0.1875mm$ is used, which corresponds to a mesh number of 1,894,892. Reflective wall boundary conditions are implemented to the top and bottom boundaries. The left and right ones are zero gradient boundaries. 

\begin{figure}[htbp]
    \centering
    \subfigure[$t=247\mu s$]
    {
        \centering
        \includegraphics[width=0.475\textwidth]{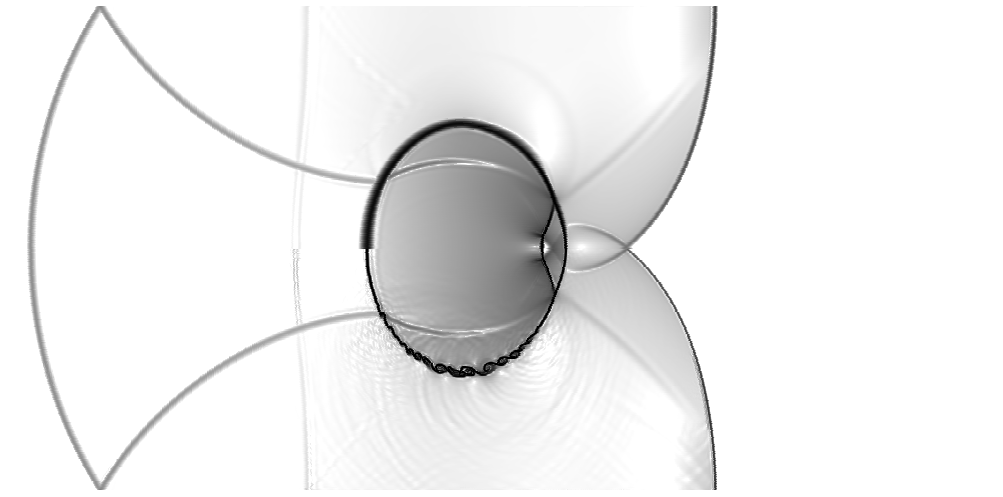}
    }
    \subfigure[$t=247\mu s$]
    {
        \centering
        \includegraphics[width=0.475\textwidth]{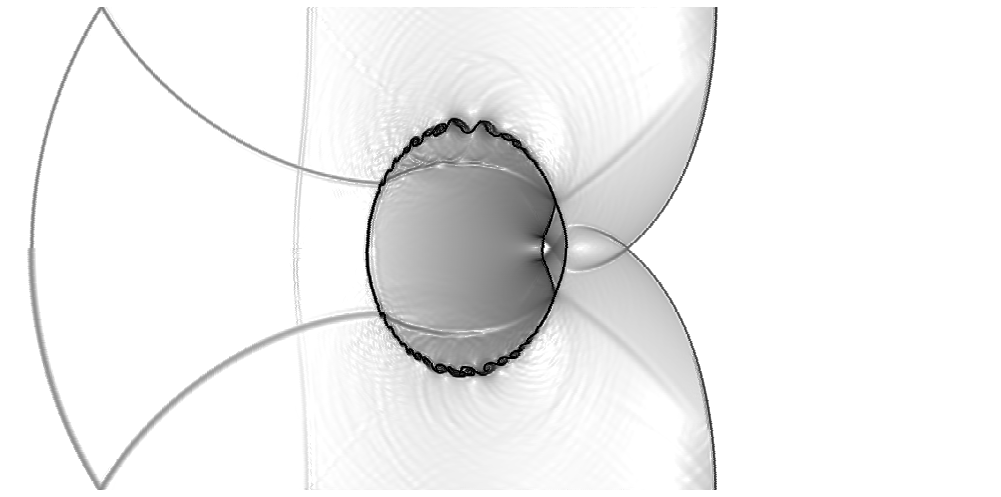}
    }
    \subfigure[$t=307\mu s$]
    {
        \centering
        \includegraphics[width=0.475\textwidth]{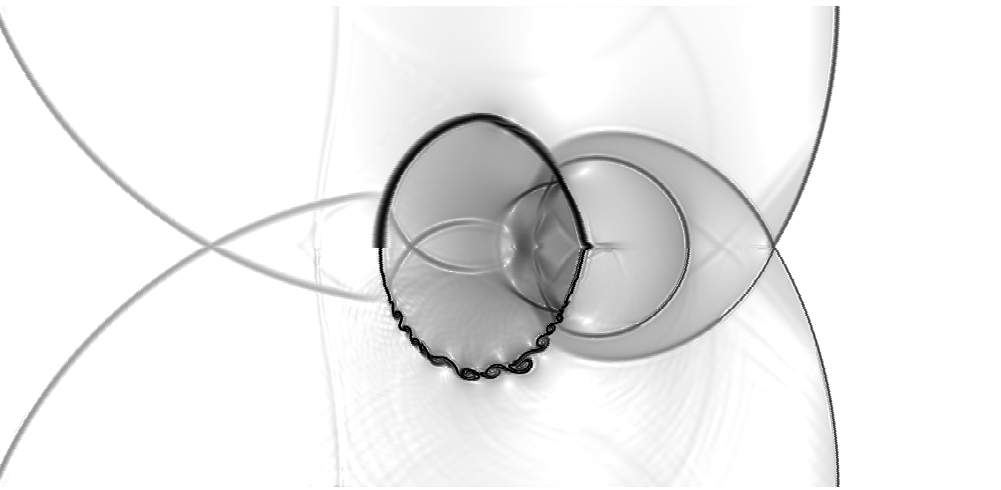}
    }
    \subfigure[$t=307\mu s$]
    {
        \centering
        \includegraphics[width=0.475\textwidth]{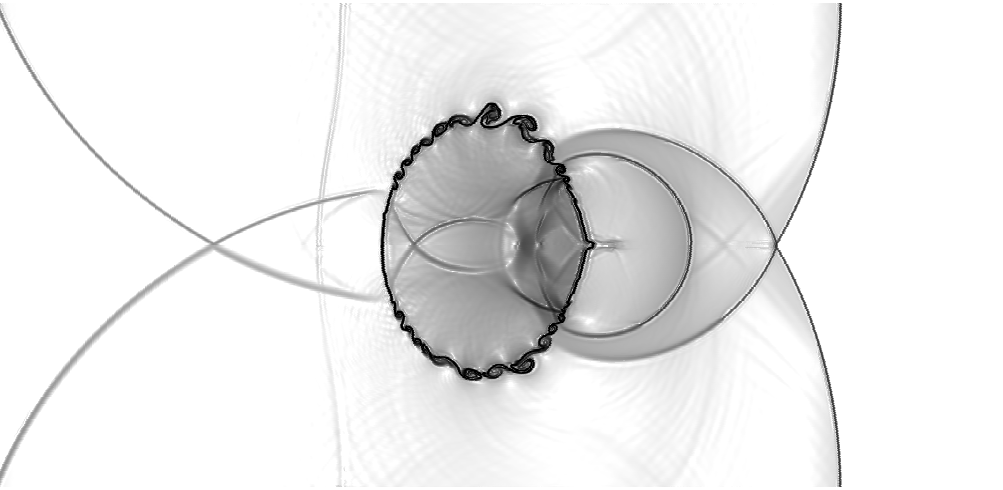}
    }
    \caption{Numerical schlierens for results of interaction between air shock and R22 cylinder. Pictures of each row are at the same instant. The result of the one-stage BVD (lower half) is plotted against results of MUSCL (left-upper half) and the two-stage BVD (right-upper half).}
    \label{fig:t6:numsh1}
\end{figure}

\begin{figure}[htbp]
    \centering
    \subfigure[$t=477\mu s$]
    {
        \centering
        \includegraphics[width=0.475\textwidth]{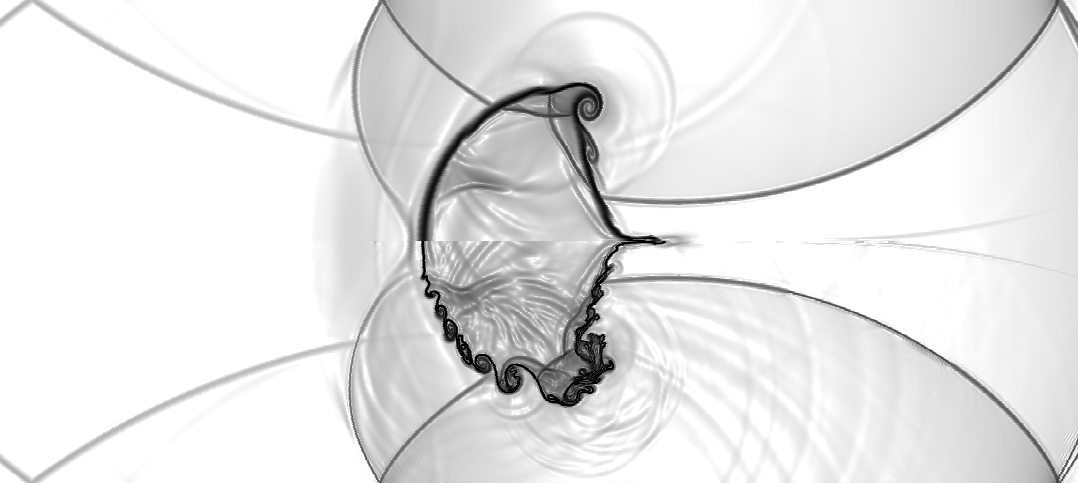}
    }
    \subfigure[$t=477\mu s$]
    {
        \centering
        \includegraphics[width=0.475\textwidth]{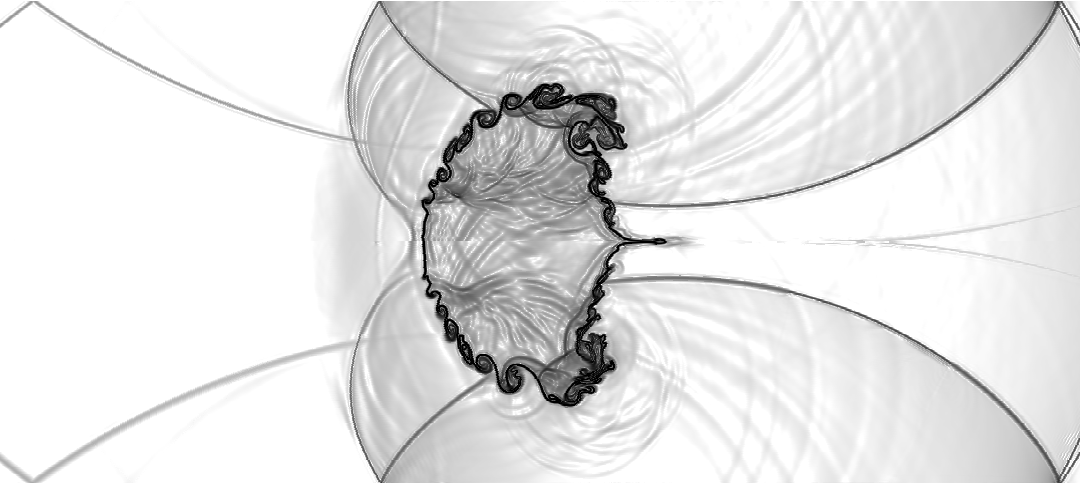}
    }

    \subfigure[$t=583\mu s$]
    {
        \centering
        \includegraphics[width=0.475\textwidth]{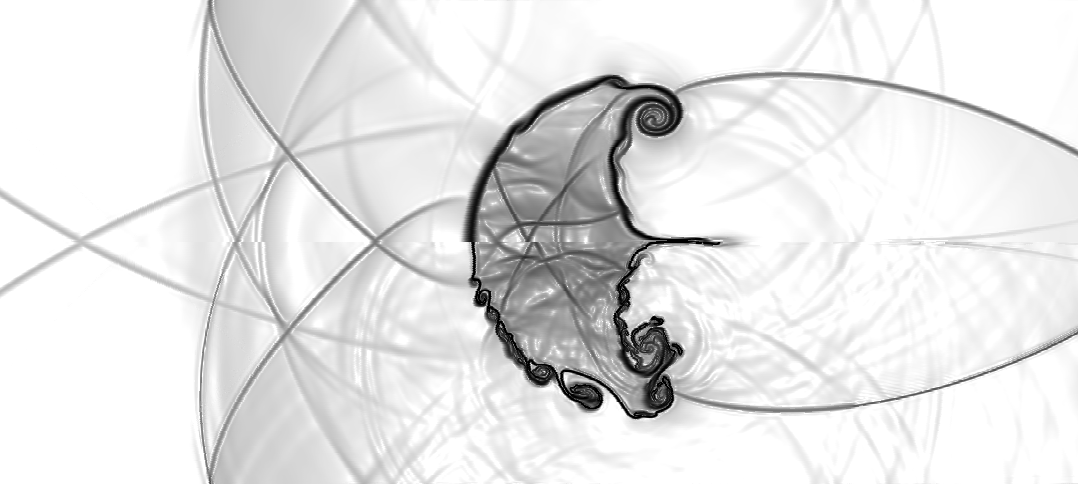}
    }
    \subfigure[$t=583\mu s$]
    {
        \centering
        \includegraphics[width=0.475\textwidth]{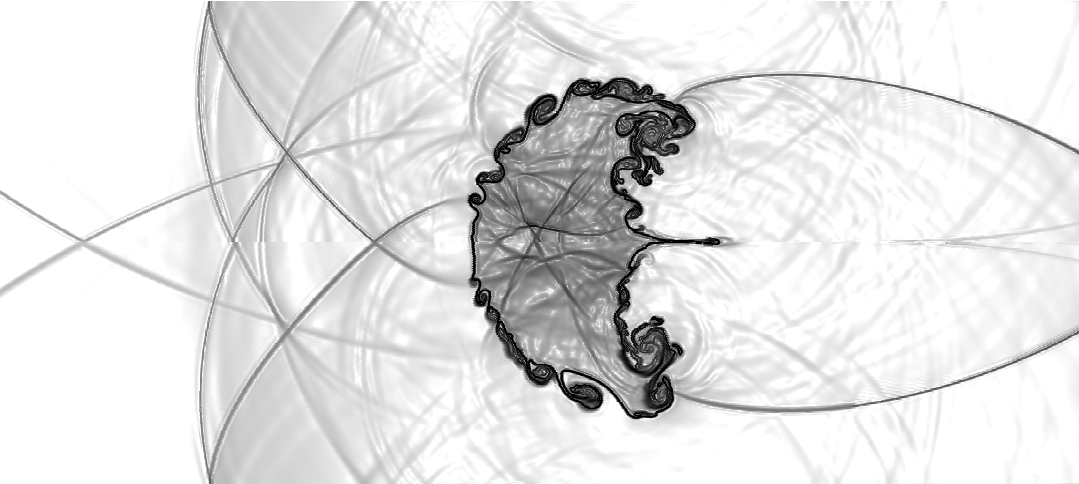}
    }
    \caption{Continue of Figure \ref{fig:t6:numsh1}}
    \label{fig:t6:numsh2}
\end{figure}

Numerical schlierens of results, $\ln(1.0+ \left| \nabla \rho \right|)$, are shown in Figure \ref{fig:t6:numsh1} and \ref{fig:t6:numsh2}. From the left column we can see that the one-stage BVD scheme can not only keep a very sharp interface, but also resolve vortices generated by the baroclinic mechanism. On the other hand, the MUSCL scheme is too diffusive that it smears almost all the small vortices. Only large-scale vortices can be observed in upper half of Figure \ref{fig:t6:numsh2} (a)(c)(e). Since Mach number of the shock wave is only 1.22, most wave structures generated by reflection and transmission are not strong enough to be distinguished by the one-stage BVD scheme. However, they can be recognised by the two-stage BVD scheme with both $\beta_l$ and $\beta_s$. At the same time, it is possible to reduce numerical dissipation further by the participation of the THINC/QQ scheme with $\beta_s$. That's why interface of results by the two-stage BVD scheme have more small structures than those from the one-stage BVD scheme.  Flow structures inside the R22 gas cylinder by BVD schemes are also clearer than the MUSCL scheme.

\begin{figure}[htbp]
    \centering
    \subfigure[$t=378\mu s$]
    {
        \centering
        \includegraphics[width=0.4\textwidth]{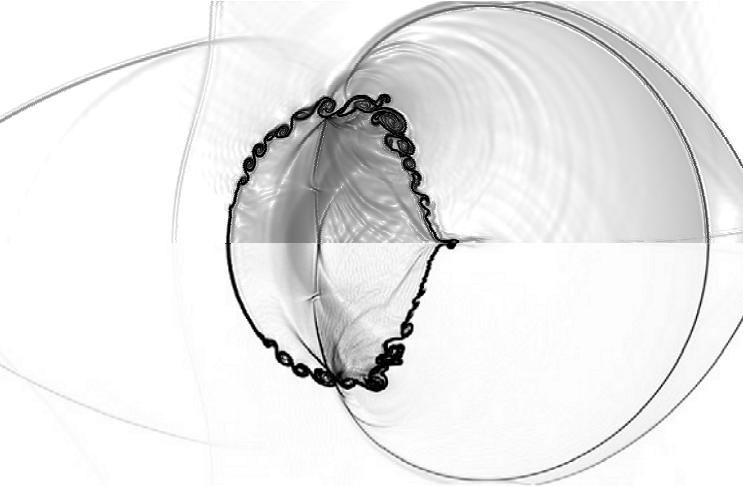}
    }
    \subfigure[$t=648\mu s$]
    {
        \centering
        \includegraphics[width=0.4\textwidth]{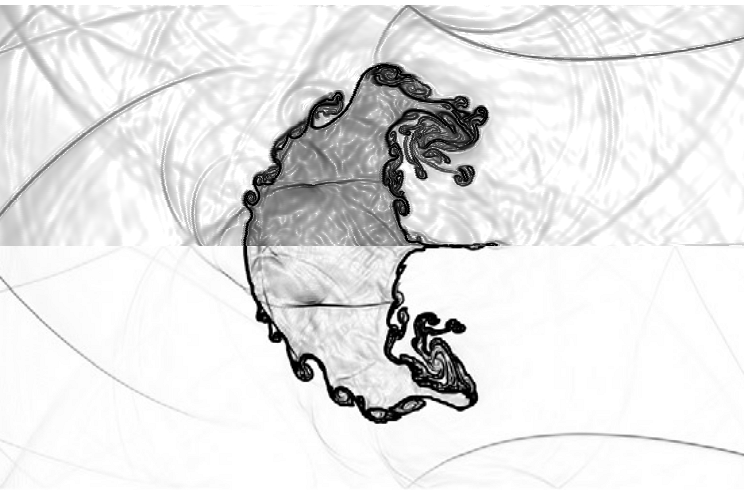}
    }
    \caption{Results from the two-stage BVD scheme (upper half) and \cite{deng2018high} (lower half) with the same mesh number.}
    \label{fig:t6:numsh3}
\end{figure}

Finally we compare our results with those from the structured BVD scheme in \cite{deng2018high}. Structured-grid framework can be done dimension-by-dimension without geometry dependence. For unstructured grids, we consider all the faces at the same time and including the geometry information of faces. Although there are some difference in detail, BVD schemes for two kind of grids share the same basic idea, say, reducing numerical diffusion by boundary variation diminishing. Thus, at least in our numerical tests, as showed in  Figure \ref{fig:t6:numsh3}, their behavior are very similar.


\section{Conclusion remarks}

In this work, we extend BVD algorithms of structured grids to unstructured grids by including geometrical information and propose two less-dissipate MUSCL-THINC/QQ-BVD schemes. Both of them can be implemented to reconstructed variables independently. Thus, they can be treated in the same manner as normal finite volume schemes. Results shows that our BVD algorithm can preserve solution properties and reduce numerical dissipation effectively. For advection or interface tracking problem, BVD schemes can limit discontinuities into few cells which can be controlled by parameter of steepness. In gas dynamic problems, both of the one- and two- stage BVD scheme can resolve sharp shock waves yet the MUSCL scheme tends to smear them. Numerical dissipation of the two-stage BVD scheme is further reduced at vortices and weak shocks. For compressible multi-component flows, the MUSCL scheme can not keep a clear material interface and resolve small flow structures induced by instabilities, but BVD schemes can capture them very clearly. Our results are comparable to results by the MUSCL-THINC-BVD scheme for structured grids. Furthermore, some of our results are even better than high-order finite volume schemes such as the fourth-order WENO scheme. This work shows that the basic idea of the BVD algorithm works very well for both structured and unstructured grids. It can be seen as another path to construct high-order non-oscillatory numerical schemes.

\section{Acknowledgment}

The authors acknowledge Dr. Peng Jin and Dr. Siengdy Tann at Tokyo Institute of Technology for their inspirational discussions. This paper was funded by International Graduate Exchange Program of Beijing Institute of Technology and National Natural Science Foundation of China (grant no. 11802178). FX was supported in part by the fund from JSPS (Japan Society for the Promotion of Science) under Grant Nos. 17K18838 and 18H01366. 

\clearpage{}
\bibliographystyle{elsarticle-num-names.bst}
\bibliography{bvdScheme} 

\begin{thebibliography}{58}
\expandafter\ifx\csname natexlab\endcsname\relax\def\natexlab#1{#1}\fi
\providecommand{\url}[1]{\texttt{#1}}
\providecommand{\href}[2]{#2}
\providecommand{\path}[1]{#1}
\providecommand{\DOIprefix}{doi:}
\providecommand{\ArXivprefix}{arXiv:}
\providecommand{\URLprefix}{URL: }
\providecommand{\Pubmedprefix}{pmid:}
\providecommand{\doi}[1]{\href{http://dx.doi.org/#1}{\path{#1}}}
\providecommand{\Pubmed}[1]{\href{pmid:#1}{\path{#1}}}
\providecommand{\bibinfo}[2]{#2}
\ifx\xfnm\relax \def\xfnm[#1]{\unskip,\space#1}\fi
\bibitem[{Barth and Frederickson(1990)}]{k1}
\bibinfo{author}{T.~Barth}, \bibinfo{author}{P.~Frederickson},
\newblock \bibinfo{title}{Higher order solution of the euler equations on
  unstructured grids using quadratic reconstruction},
\newblock in: \bibinfo{booktitle}{28th aerospace sciences meeting},
  \bibinfo{year}{1990}, p.~\bibinfo{pages}{13}.
\bibitem[{BARTH(1993)}]{k2}
\bibinfo{author}{T.~BARTH},
\newblock \bibinfo{title}{Recent developments in high order k-exact
  reconstruction on unstructured meshes},
\newblock in: \bibinfo{booktitle}{31st Aerospace Sciences Meeting},
  \bibinfo{year}{1993}, p. \bibinfo{pages}{668}.
\bibitem[{Delanaye and Liu(1999)}]{k3}
\bibinfo{author}{M.~Delanaye}, \bibinfo{author}{Y.~Liu},
\newblock \bibinfo{title}{Quadratic reconstruction finite volume schemes on 3d
  arbitrary unstructured polyhedral grids},
\newblock in: \bibinfo{booktitle}{14th Computational Fluid Dynamics
  Conference}, \bibinfo{year}{1999}, p. \bibinfo{pages}{3259}.
\bibitem[{Reed and Hill(????)}]{dg1}
\bibinfo{author}{W.~Reed}, \bibinfo{author}{T.~Hill},
\newblock \bibinfo{title}{Triangular mesh methods for the neutron transport
  equation, technical report la-ur-73-479, los alamos scientific laboratory,
  los alamos, 1973}  (????).
\bibitem[{Cockburn et~al.(1990)Cockburn, Hou, and Shu}]{dg2}
\bibinfo{author}{B.~Cockburn}, \bibinfo{author}{S.~Hou}, \bibinfo{author}{C.-W.
  Shu},
\newblock \bibinfo{title}{The runge-kutta local projection discontinuous
  galerkin finite element method for conservation laws. iv. the
  multidimensional case},
\newblock \bibinfo{journal}{Mathematics of Computation} \bibinfo{volume}{54}
  (\bibinfo{year}{1990}) \bibinfo{pages}{545--581}.
\bibitem[{Huynh(2007)}]{fr1}
\bibinfo{author}{H.~T. Huynh},
\newblock \bibinfo{title}{A flux reconstruction approach to high-order schemes
  including discontinuous galerkin methods},
\newblock in: \bibinfo{booktitle}{18th AIAA Computational Fluid Dynamics
  Conference}, \bibinfo{year}{2007}, p. \bibinfo{pages}{4079}.
\bibitem[{Vincent et~al.(2011)Vincent, Castonguay, and Jameson}]{fr2}
\bibinfo{author}{P.~E. Vincent}, \bibinfo{author}{P.~Castonguay},
  \bibinfo{author}{A.~Jameson},
\newblock \bibinfo{title}{A new class of high-order energy stable flux
  reconstruction schemes},
\newblock \bibinfo{journal}{Journal of Scientific Computing}
  \bibinfo{volume}{47} (\bibinfo{year}{2011}) \bibinfo{pages}{50--72}.
\bibitem[{Castonguay et~al.(2012)Castonguay, Vincent, and Jameson}]{fr3}
\bibinfo{author}{P.~Castonguay}, \bibinfo{author}{P.~E. Vincent},
  \bibinfo{author}{A.~Jameson},
\newblock \bibinfo{title}{A new class of high-order energy stable flux
  reconstruction schemes for triangular elements},
\newblock \bibinfo{journal}{Journal of Scientific Computing}
  \bibinfo{volume}{51} (\bibinfo{year}{2012}) \bibinfo{pages}{224--256}.
\bibitem[{Witherden et~al.(2014)Witherden, Farrington, and Vincent}]{fr4}
\bibinfo{author}{F.~D. Witherden}, \bibinfo{author}{A.~M. Farrington},
  \bibinfo{author}{P.~E. Vincent},
\newblock \bibinfo{title}{Pyfr: An open source framework for solving
  advection--diffusion type problems on streaming architectures using the flux
  reconstruction approach},
\newblock \bibinfo{journal}{Computer Physics Communications}
  \bibinfo{volume}{185} (\bibinfo{year}{2014}) \bibinfo{pages}{3028--3040}.
\bibitem[{Kopriva and Kolias(1996)}]{sd1}
\bibinfo{author}{D.~A. Kopriva}, \bibinfo{author}{J.~H. Kolias},
\newblock \bibinfo{title}{A conservative staggered-grid chebyshev multidomain
  method for compressible flows},
\newblock \bibinfo{journal}{Journal of computational physics}
  \bibinfo{volume}{125} (\bibinfo{year}{1996}) \bibinfo{pages}{244--261}.
\bibitem[{Sun et~al.(2006)Sun, Wang, and Liu}]{sd2}
\bibinfo{author}{Y.~Sun}, \bibinfo{author}{Z.~Wang}, \bibinfo{author}{Y.~Liu},
\newblock \bibinfo{title}{High-order multidomain spectral difference method for
  the navier-stokes equations},
\newblock in: \bibinfo{booktitle}{44th AIAA Aerospace Sciences Meeting and
  Exhibit}, \bibinfo{year}{2006}, p. \bibinfo{pages}{301}.
\bibitem[{Harten(1983)}]{harten1983high}
\bibinfo{author}{A.~Harten},
\newblock \bibinfo{title}{High resolution schemes for hyperbolic conservation
  laws},
\newblock \bibinfo{journal}{Journal of computational physics}
  \bibinfo{volume}{49} (\bibinfo{year}{1983}) \bibinfo{pages}{357--393}.
\bibitem[{Barth and Jespersen(1989)}]{tvdm1}
\bibinfo{author}{T.~Barth}, \bibinfo{author}{D.~Jespersen},
\newblock \bibinfo{title}{The design and application of upwind schemes on
  unstructured meshes},
\newblock in: \bibinfo{booktitle}{27th Aerospace sciences meeting},
  \bibinfo{year}{1989}, p. \bibinfo{pages}{366}.
\bibitem[{Spekreijse(1987)}]{tvdm2}
\bibinfo{author}{S.~Spekreijse},
\newblock \bibinfo{title}{Multigrid solution of monotone second-order
  discretizations of hyperbolic conservation laws},
\newblock \bibinfo{journal}{Mathematics of Computation} \bibinfo{volume}{49}
  (\bibinfo{year}{1987}) \bibinfo{pages}{135--155}.
\bibitem[{Park et~al.(2010)Park, Yoon, and Kim}]{mlp1}
\bibinfo{author}{J.~S. Park}, \bibinfo{author}{S.-H. Yoon},
  \bibinfo{author}{C.~Kim},
\newblock \bibinfo{title}{Multi-dimensional limiting process for hyperbolic
  conservation laws on unstructured grids},
\newblock \bibinfo{journal}{Journal of Computational Physics}
  \bibinfo{volume}{229} (\bibinfo{year}{2010}) \bibinfo{pages}{788--812}.
\bibitem[{Park and Kim(2012)}]{park2012multi}
\bibinfo{author}{J.~S. Park}, \bibinfo{author}{C.~Kim},
\newblock \bibinfo{title}{Multi-dimensional limiting process for finite volume
  methods on unstructured grids},
\newblock \bibinfo{journal}{Computers \& Fluids} \bibinfo{volume}{65}
  (\bibinfo{year}{2012}) \bibinfo{pages}{8--24}.
\bibitem[{Friedrich(1998)}]{friedrich1998weighted}
\bibinfo{author}{O.~Friedrich},
\newblock \bibinfo{title}{Weighted essentially non-oscillatory schemes for the
  interpolation of mean values on unstructured grids},
\newblock \bibinfo{journal}{Journal of computational physics}
  \bibinfo{volume}{144} (\bibinfo{year}{1998}) \bibinfo{pages}{194--212}.
\bibitem[{Haselbacher(2005)}]{haselbacher2005weno}
\bibinfo{author}{A.~Haselbacher},
\newblock \bibinfo{title}{A weno reconstruction algorithim for unstructured
  grids based on explicit stencil construction},
\newblock in: \bibinfo{booktitle}{43rd AIAA Aerospace Sciences Meeting and
  Exhibit}, \bibinfo{year}{2005}, p. \bibinfo{pages}{879}.
\bibitem[{Wolf and Azevedo(2007)}]{wolf2007high}
\bibinfo{author}{W.~Wolf}, \bibinfo{author}{J.~Azevedo},
\newblock \bibinfo{title}{High-order eno and weno schemes for unstructured
  grids},
\newblock \bibinfo{journal}{International Journal for Numerical Methods in
  Fluids} \bibinfo{volume}{55} (\bibinfo{year}{2007})
  \bibinfo{pages}{917--943}.
\bibitem[{Titarev et~al.(2010)Titarev, Tsoutsanis, and
  Drikakis}]{titarev2010weno}
\bibinfo{author}{V.~Titarev}, \bibinfo{author}{P.~Tsoutsanis},
  \bibinfo{author}{D.~Drikakis},
\newblock \bibinfo{title}{Weno schemes for mixed-element unstructured meshes},
\newblock \bibinfo{journal}{Communications in Computational Physics}
  \bibinfo{volume}{8} (\bibinfo{year}{2010}) \bibinfo{pages}{585}.
\bibitem[{Tsoutsanis et~al.(2011)Tsoutsanis, Titarev, and
  Drikakis}]{tsoutsanis2011weno}
\bibinfo{author}{P.~Tsoutsanis}, \bibinfo{author}{V.~A. Titarev},
  \bibinfo{author}{D.~Drikakis},
\newblock \bibinfo{title}{Weno schemes on arbitrary mixed-element unstructured
  meshes in three space dimensions},
\newblock \bibinfo{journal}{Journal of Computational Physics}
  \bibinfo{volume}{230} (\bibinfo{year}{2011}) \bibinfo{pages}{1585--1601}.
\bibitem[{Persson and Peraire(2006)}]{av1}
\bibinfo{author}{P.-O. Persson}, \bibinfo{author}{J.~Peraire},
\newblock \bibinfo{title}{Sub-cell shock capturing for discontinuous galerkin
  methods},
\newblock in: \bibinfo{booktitle}{44th AIAA Aerospace Sciences Meeting and
  Exhibit}, \bibinfo{year}{2006}, p. \bibinfo{pages}{112}.
\bibitem[{Barter and Darmofal(2010)}]{av2}
\bibinfo{author}{G.~E. Barter}, \bibinfo{author}{D.~L. Darmofal},
\newblock \bibinfo{title}{Shock capturing with pde-based artificial viscosity
  for dgfem: Part i. formulation},
\newblock \bibinfo{journal}{Journal of Computational Physics}
  \bibinfo{volume}{229} (\bibinfo{year}{2010}) \bibinfo{pages}{1810--1827}.
\bibitem[{Dumbser and Loub{\`e}re(2016)}]{sb1}
\bibinfo{author}{M.~Dumbser}, \bibinfo{author}{R.~Loub{\`e}re},
\newblock \bibinfo{title}{A simple robust and accurate a posteriori sub-cell
  finite volume limiter for the discontinuous galerkin method on unstructured
  meshes},
\newblock \bibinfo{journal}{Journal of Computational Physics}
  \bibinfo{volume}{319} (\bibinfo{year}{2016}) \bibinfo{pages}{163--199}.
\bibitem[{Dumbser et~al.(2014)Dumbser, Zanotti, Loub{\`e}re, and Diot}]{sb2}
\bibinfo{author}{M.~Dumbser}, \bibinfo{author}{O.~Zanotti},
  \bibinfo{author}{R.~Loub{\`e}re}, \bibinfo{author}{S.~Diot},
\newblock \bibinfo{title}{A posteriori subcell limiting of the discontinuous
  galerkin finite element method for hyperbolic conservation laws},
\newblock \bibinfo{journal}{Journal of Computational Physics}
  \bibinfo{volume}{278} (\bibinfo{year}{2014}) \bibinfo{pages}{47--75}.
\bibitem[{Sun et~al.(2016)Sun, Inaba, and Xiao}]{sun2016boundary}
\bibinfo{author}{Z.~Sun}, \bibinfo{author}{S.~Inaba},
  \bibinfo{author}{F.~Xiao},
\newblock \bibinfo{title}{Boundary variation diminishing (bvd) reconstruction:
  A new approach to improve godunov schemes},
\newblock \bibinfo{journal}{Journal of Computational Physics}
  \bibinfo{volume}{322} (\bibinfo{year}{2016}) \bibinfo{pages}{309--325}.
\bibitem[{Xiao et~al.(2005)Xiao, Honma, and Kono}]{xiao2005simple}
\bibinfo{author}{F.~Xiao}, \bibinfo{author}{Y.~Honma},
  \bibinfo{author}{T.~Kono},
\newblock \bibinfo{title}{A simple algebraic interface capturing scheme using
  hyperbolic tangent function},
\newblock \bibinfo{journal}{International Journal for Numerical Methods in
  Fluids} \bibinfo{volume}{48} (\bibinfo{year}{2005})
  \bibinfo{pages}{1023--1040}.
\bibitem[{Jiang and Shu(1996)}]{jiang1996efficient}
\bibinfo{author}{G.-S. Jiang}, \bibinfo{author}{C.-W. Shu},
\newblock \bibinfo{title}{Efficient implementation of weighted eno schemes},
\newblock \bibinfo{journal}{Journal of computational physics}
  \bibinfo{volume}{126} (\bibinfo{year}{1996}) \bibinfo{pages}{202--228}.
\bibitem[{Deng et~al.(2018{\natexlab{a}})Deng, Xie, Loub{\`e}re, Shimizu, and
  Xiao}]{deng2018limiter}
\bibinfo{author}{X.~Deng}, \bibinfo{author}{B.~Xie},
  \bibinfo{author}{R.~Loub{\`e}re}, \bibinfo{author}{Y.~Shimizu},
  \bibinfo{author}{F.~Xiao},
\newblock \bibinfo{title}{Limiter-free discontinuity-capturing scheme for
  compressible gas dynamics with reactive fronts},
\newblock \bibinfo{journal}{Computers \& Fluids} \bibinfo{volume}{171}
  (\bibinfo{year}{2018}{\natexlab{a}}) \bibinfo{pages}{1--14}.
\bibitem[{Deng et~al.(2018{\natexlab{b}})Deng, Inaba, Xie, Shyue, and
  Xiao}]{deng2018high}
\bibinfo{author}{X.~Deng}, \bibinfo{author}{S.~Inaba},
  \bibinfo{author}{B.~Xie}, \bibinfo{author}{K.-M. Shyue},
  \bibinfo{author}{F.~Xiao},
\newblock \bibinfo{title}{High fidelity discontinuity-resolving reconstruction
  for compressible multiphase flows with moving interfaces},
\newblock \bibinfo{journal}{Journal of Computational Physics}
  \bibinfo{volume}{371} (\bibinfo{year}{2018}{\natexlab{b}})
  \bibinfo{pages}{945--966}.
\bibitem[{Deng et~al.(2019)Deng, Shimizu, and Xiao}]{deng2019fifth}
\bibinfo{author}{X.~Deng}, \bibinfo{author}{Y.~Shimizu},
  \bibinfo{author}{F.~Xiao},
\newblock \bibinfo{title}{A fifth-order shock capturing scheme with two-stage
  boundary variation diminishing algorithm},
\newblock \bibinfo{journal}{Journal of Computational Physics}
  \bibinfo{volume}{386} (\bibinfo{year}{2019}) \bibinfo{pages}{323--349}.
\bibitem[{Deng et~al.(2020)Deng, Shimizu, Xie, and Xiao}]{deng2020constructing}
\bibinfo{author}{X.~Deng}, \bibinfo{author}{Y.~Shimizu},
  \bibinfo{author}{B.~Xie}, \bibinfo{author}{F.~Xiao},
\newblock \bibinfo{title}{Constructing higher order discontinuity-capturing
  schemes with upwind-biased interpolations and boundary variation diminishing
  algorithm},
\newblock \bibinfo{journal}{Computers \& Fluids}  (\bibinfo{year}{2020})
  \bibinfo{pages}{104433}.
\bibitem[{Xie and Xiao(2017)}]{xie2017toward}
\bibinfo{author}{B.~Xie}, \bibinfo{author}{F.~Xiao},
\newblock \bibinfo{title}{Toward efficient and accurate interface capturing on
  arbitrary hybrid unstructured grids: The thinc method with quadratic surface
  representation and gaussian quadrature},
\newblock \bibinfo{journal}{Journal of Computational Physics}
  \bibinfo{volume}{349} (\bibinfo{year}{2017}) \bibinfo{pages}{415--440}.
\bibitem[{Xie et~al.(2019)Xie, Deng, Liao, and Xiao}]{xie2019high}
\bibinfo{author}{B.~Xie}, \bibinfo{author}{X.~Deng}, \bibinfo{author}{S.~Liao},
  \bibinfo{author}{F.~Xiao},
\newblock \bibinfo{title}{High-order multi-moment finite volume method with
  smoothness adaptive fitting reconstruction for compressible viscous flow},
\newblock \bibinfo{journal}{Journal of Computational Physics}
  \bibinfo{volume}{394} (\bibinfo{year}{2019}) \bibinfo{pages}{559 -- 593}.
\bibitem[{Allaire et~al.(2002)Allaire, Clerc, and Kokh}]{allaire2002five}
\bibinfo{author}{G.~Allaire}, \bibinfo{author}{S.~Clerc},
  \bibinfo{author}{S.~Kokh},
\newblock \bibinfo{title}{A five-equation model for the simulation of
  interfaces between compressible fluids},
\newblock \bibinfo{journal}{Journal of Computational Physics}
  \bibinfo{volume}{181} (\bibinfo{year}{2002}) \bibinfo{pages}{577--616}.
\bibitem[{Shyue(2001)}]{shyue2001fluid}
\bibinfo{author}{K.-M. Shyue},
\newblock \bibinfo{title}{A fluid-mixture type algorithm for compressible
  multicomponent flow with mie--gr{\"u}neisen equation of state},
\newblock \bibinfo{journal}{Journal of Computational Physics}
  \bibinfo{volume}{171} (\bibinfo{year}{2001}) \bibinfo{pages}{678--707}.
\bibitem[{Gottlieb et~al.(2001)Gottlieb, Shu, and Tadmor}]{gottlieb2001strong}
\bibinfo{author}{S.~Gottlieb}, \bibinfo{author}{C.-W. Shu},
  \bibinfo{author}{E.~Tadmor},
\newblock \bibinfo{title}{Strong stability-preserving high-order time
  discretization methods},
\newblock \bibinfo{journal}{SIAM review} \bibinfo{volume}{43}
  (\bibinfo{year}{2001}) \bibinfo{pages}{89--112}.
\bibitem[{Johnsen and Colonius(2006)}]{johnsen2006implementation}
\bibinfo{author}{E.~Johnsen}, \bibinfo{author}{T.~Colonius},
\newblock \bibinfo{title}{Implementation of weno schemes in compressible
  multicomponent flow problems},
\newblock \bibinfo{journal}{Journal of Computational Physics}
  \bibinfo{volume}{219} (\bibinfo{year}{2006}) \bibinfo{pages}{715--732}.
\bibitem[{Johnsen(2011)}]{johnsen20118665}
\bibinfo{author}{E.~Johnsen},
\newblock \bibinfo{title}{On the treatment of contact discontinuities using
  weno schemes},
\newblock \bibinfo{journal}{Journal of Computational Physics}
  \bibinfo{volume}{230} (\bibinfo{year}{2011}) \bibinfo{pages}{8665 -- 8668}.
\bibitem[{Harten et~al.(1983)Harten, Lax, and Leer}]{harten1983upstream}
\bibinfo{author}{A.~Harten}, \bibinfo{author}{P.~D. Lax},
  \bibinfo{author}{B.~v. Leer},
\newblock \bibinfo{title}{On upstream differencing and godunov-type schemes for
  hyperbolic conservation laws},
\newblock \bibinfo{journal}{SIAM review} \bibinfo{volume}{25}
  (\bibinfo{year}{1983}) \bibinfo{pages}{35--61}.
\bibitem[{Tann et~al.(2019)Tann, Deng, Shimizu, Loub{\`e}re, and
  Xiao}]{tann2019solution}
\bibinfo{author}{S.~Tann}, \bibinfo{author}{X.~Deng},
  \bibinfo{author}{Y.~Shimizu}, \bibinfo{author}{R.~Loub{\`e}re},
  \bibinfo{author}{F.~Xiao},
\newblock \bibinfo{title}{Solution property preserving reconstruction for
  finite volume scheme: a bvd+ mood framework},
\newblock \bibinfo{journal}{International Journal for Numerical Methods in
  Fluids}  (\bibinfo{year}{2019}).
\bibitem[{LeVeque(1996)}]{leveque1996high}
\bibinfo{author}{R.~J. LeVeque},
\newblock \bibinfo{title}{High-resolution conservative algorithms for advection
  in incompressible flow},
\newblock \bibinfo{journal}{SIAM Journal on Numerical Analysis}
  \bibinfo{volume}{33} (\bibinfo{year}{1996}) \bibinfo{pages}{627--665}.
\bibitem[{Zalesak(1979)}]{zalesak1979fully}
\bibinfo{author}{S.~T. Zalesak},
\newblock \bibinfo{title}{Fully multidimensional flux-corrected transport
  algorithms for fluids},
\newblock \bibinfo{journal}{Journal of computational physics}
  \bibinfo{volume}{31} (\bibinfo{year}{1979}) \bibinfo{pages}{335--362}.
\bibitem[{Toro(2013)}]{toro2013riemann}
\bibinfo{author}{E.~F. Toro}, \bibinfo{title}{Riemann solvers and numerical
  methods for fluid dynamics: a practical introduction},
  \bibinfo{publisher}{Springer Science \& Business Media},
  \bibinfo{year}{2013}.
\bibitem[{Woodward and Colella(1984)}]{woodward1984numerical}
\bibinfo{author}{P.~Woodward}, \bibinfo{author}{P.~Colella},
\newblock \bibinfo{title}{The numerical simulation of two-dimensional fluid
  flow with strong shocks},
\newblock \bibinfo{journal}{Journal of computational physics}
  \bibinfo{volume}{54} (\bibinfo{year}{1984}) \bibinfo{pages}{115--173}.
\bibitem[{Hu and Shu(1999)}]{hu1999weighted}
\bibinfo{author}{C.~Hu}, \bibinfo{author}{C.-W. Shu},
\newblock \bibinfo{title}{Weighted essentially non-oscillatory schemes on
  triangular meshes},
\newblock \bibinfo{journal}{Journal of Computational Physics}
  \bibinfo{volume}{150} (\bibinfo{year}{1999}) \bibinfo{pages}{97--127}.
\bibitem[{Li and Ren(2012)}]{li2012high}
\bibinfo{author}{W.~Li}, \bibinfo{author}{Y.-X. Ren},
\newblock \bibinfo{title}{High-order k-exact weno finite volume schemes for
  solving gas dynamic euler equations on unstructured grids},
\newblock \bibinfo{journal}{International Journal for Numerical Methods in
  Fluids} \bibinfo{volume}{70} (\bibinfo{year}{2012})
  \bibinfo{pages}{742--763}.
\bibitem[{Zhu et~al.(2008)Zhu, Qiu, Shu, and Dumbser}]{zhu2008runge}
\bibinfo{author}{J.~Zhu}, \bibinfo{author}{J.~Qiu}, \bibinfo{author}{C.-W.
  Shu}, \bibinfo{author}{M.~Dumbser},
\newblock \bibinfo{title}{Runge--kutta discontinuous galerkin method using weno
  limiters ii: unstructured meshes},
\newblock \bibinfo{journal}{Journal of Computational Physics}
  \bibinfo{volume}{227} (\bibinfo{year}{2008}) \bibinfo{pages}{4330--4353}.
\bibitem[{Zhu and Qiu(2009)}]{zhu2009hermite}
\bibinfo{author}{J.~Zhu}, \bibinfo{author}{J.~Qiu},
\newblock \bibinfo{title}{Hermite weno schemes and their application as
  limiters for runge-kutta discontinuous galerkin method, iii: unstructured
  meshes},
\newblock \bibinfo{journal}{Journal of Scientific Computing}
  \bibinfo{volume}{39} (\bibinfo{year}{2009}) \bibinfo{pages}{293--321}.
\bibitem[{Shyue and Xiao(2014)}]{shyue2014eulerian}
\bibinfo{author}{K.-M. Shyue}, \bibinfo{author}{F.~Xiao},
\newblock \bibinfo{title}{An eulerian interface sharpening algorithm for
  compressible two-phase flow: the algebraic thinc approach},
\newblock \bibinfo{journal}{Journal of Computational Physics}
  \bibinfo{volume}{268} (\bibinfo{year}{2014}) \bibinfo{pages}{326--354}.
\bibitem[{Pandare et~al.(2019)Pandare, Luo, and Bakosi}]{pandare2019enhanced}
\bibinfo{author}{A.~K. Pandare}, \bibinfo{author}{H.~Luo},
  \bibinfo{author}{J.~Bakosi},
\newblock \bibinfo{title}{An enhanced ausm+up scheme for high-speed
  compressible two-phase flows on hybrid grids},
\newblock \bibinfo{journal}{Shock Waves} \bibinfo{volume}{29}
  (\bibinfo{year}{2019}) \bibinfo{pages}{629--649}.
\bibitem[{Quirk and Karni(1996)}]{quirk1996dynamics}
\bibinfo{author}{J.~J. Quirk}, \bibinfo{author}{S.~Karni},
\newblock \bibinfo{title}{On the dynamics of a shock--bubble interaction},
\newblock \bibinfo{journal}{Journal of Fluid Mechanics} \bibinfo{volume}{318}
  (\bibinfo{year}{1996}) \bibinfo{pages}{129--163}.
\bibitem[{Shyue(2006)}]{shyue2006wave}
\bibinfo{author}{K.-M. Shyue},
\newblock \bibinfo{title}{A wave-propagation based volume tracking method for
  compressible multicomponent flow in two space dimensions},
\newblock \bibinfo{journal}{Journal of Computational Physics}
  \bibinfo{volume}{215} (\bibinfo{year}{2006}) \bibinfo{pages}{219--244}.
\bibitem[{Shankar et~al.(2010)Shankar, Kawai, and Lele}]{shankar2010numerical}
\bibinfo{author}{S.~Shankar}, \bibinfo{author}{S.~Kawai},
  \bibinfo{author}{S.~Lele},
\newblock \bibinfo{title}{Numerical simulation of multicomponent shock
  accelerated flows and mixing using localized artificial diffusivity method},
\newblock in: \bibinfo{booktitle}{48th AIAA Aerospace Sciences Meeting
  Including the New Horizons Forum and Aerospace Exposition},
  \bibinfo{year}{2010}, p. \bibinfo{pages}{352}.
\bibitem[{So et~al.(2012)So, Hu, and Adams}]{so2012anti}
\bibinfo{author}{K.~So}, \bibinfo{author}{X.~Hu}, \bibinfo{author}{N.~A.
  Adams},
\newblock \bibinfo{title}{Anti-diffusion interface sharpening technique for
  two-phase compressible flow simulations},
\newblock \bibinfo{journal}{Journal of Computational Physics}
  \bibinfo{volume}{231} (\bibinfo{year}{2012}) \bibinfo{pages}{4304--4323}.
\bibitem[{Picone and Boris(1988)}]{picone1988vorticity}
\bibinfo{author}{J.~Picone}, \bibinfo{author}{J.~Boris},
\newblock \bibinfo{title}{Vorticity generation by shock propagation through
  bubbles in a gas},
\newblock \bibinfo{journal}{Journal of Fluid Mechanics} \bibinfo{volume}{189}
  (\bibinfo{year}{1988}) \bibinfo{pages}{23--51}.
\bibitem[{Giordano and Burtschell(2006)}]{giordano2006richtmyer}
\bibinfo{author}{J.~Giordano}, \bibinfo{author}{Y.~Burtschell},
\newblock \bibinfo{title}{Richtmyer-meshkov instability induced by shock-bubble
  interaction: Numerical and analytical studies with experimental validation},
\newblock \bibinfo{journal}{Physics of Fluids} \bibinfo{volume}{18}
  (\bibinfo{year}{2006}) \bibinfo{pages}{036102}.
\bibitem[{Haas and Sturtevant(1987)}]{haas1987interaction}
\bibinfo{author}{J.-F. Haas}, \bibinfo{author}{B.~Sturtevant},
\newblock \bibinfo{title}{Interaction of weak shock waves with cylindrical and
  spherical gas inhomogeneities},
\newblock \bibinfo{journal}{Journal of Fluid Mechanics} \bibinfo{volume}{181}
  (\bibinfo{year}{1987}) \bibinfo{pages}{41--76}.

\end{thebibliography}

\end{document}